\documentclass[11pt,oneside,notitlepage]{article}

\usepackage[letterpaper, total={6.5in, 8.5in}]{geometry}

\usepackage{amsmath}
\usepackage{amssymb}
\usepackage{amsthm}
\usepackage{array}
\usepackage{bm}
\usepackage{mathrsfs}
\usepackage{mathtools}

\usepackage[font=small]{caption} 
\usepackage[font=small]{subcaption}

\usepackage{cite}
\usepackage{comment}
\usepackage{enumitem}
\usepackage{float}
\usepackage[T1]{fontenc}
\usepackage{graphicx}
\usepackage[
colorlinks=true,
citecolor=MidnightBlue,
linkcolor=BrickRed,
urlcolor=MidnightBlue,
breaklinks=true,
linktoc=all
]{hyperref}
\usepackage{listings}
\usepackage{multicol}
\usepackage{stmaryrd}
\usepackage{textgreek}
\usepackage{titlesec}
\usepackage{tocloft}
\usepackage{wrapfig}
\usepackage[usenames,dvipsnames]{xcolor}
\usepackage{xy}
\usepackage{amsmath}
\usepackage{relsize}
\usepackage{empheq}
\usepackage{comment}
\usepackage{booktabs}
\usepackage{layouts}
\usepackage{environ}
\usepackage{pdflscape}
\usepackage{afterpage}
\usepackage{placeins} 
\usepackage[normalem]{ulem} 
\usepackage[all]{nowidow} 

\usepackage{pgfplots}
\pgfplotsset{compat=1.15}
\usepackage{pgfmath}
\usetikzlibrary{patterns, calc, shapes.geometric, positioning, arrows.meta,backgrounds}
\usepackage{tikz-imagelabels}
\usepackage{tikz}

\let\originalleft\left
\let\originalright\right
\newcommand{\leftR}{\mathopen{}\mathclose\bgroup\originalleft}
\newcommand{\rightR}{\aftergroup\egroup\originalright}

\newcommand{\diff}[1]{~\mathrm{d}#1}
\newcommand{\ddiff}[1]{\mathrm{d}#1}

\newcommand{\ChebT}[2]{P_{#1}\leftR(#2\rightR)}
\newcommand{\ChebTR}[1]{P_{#1}}

\newcommand{\divv}[1]{\mathrm{div}\leftR(#1\rightR)}
\newcommand{\grad}[1]{\mathrm{grad}\leftR(#1\rightR)}
\newcommand{\laplaces}[1]{\Delta_\mathrm{s} #1}
\DeclareMathOperator{\tr}{tr}
\newcommand{\bbar}[1]{\Bar{\Bar{#1}}}

\definecolor{myblue}{rgb}{0, 0, 0.901}%
\definecolor{myorange}{rgb}{0.901, 0.364, 0.090}%
\definecolor{mygreen}{rgb}{0.160, 0.6, 0.031}%

\newcommand{\phik}[1]{\phi_{#1} \leftR(\theta^\alpha \rightR)}

\newcommand{\textspace}{\smallskip}

\tikzset{
    position/.style args={#1:#2 from #3}{
        at=(#3), xshift=#1, yshift=#2
    }
}

\newcommand{\bma}{\bm{a}}
\newcommand{\bmb}{\bm{b}}

\newcommand{\bme}{\bm{e}}
\newcommand{\bmf}{\bm{f}}
\newcommand{\bmg}{\bm{g}}

\newcommand{\bmi}{\bm{i}}

\newcommand{\bmn}{\bm{n}}

\newcommand{\bmt}{\bm{t}}
\newcommand{\bmu}{\bm{u}}
\newcommand{\bmv}{\bm{v}}

\newcommand{\bmx}{\bm{x}}

\newcommand{\bmD}{\bm{D}}

\newcommand{\bmI}{\bm{I}}
\newcommand{\bmJ}{\bm{J}}

\newcommand{\bmT}{\bm{T}}

\newcommand{\Ga}{\alpha}
\newcommand{\Gb}{\beta}
\newcommand{\Gg}{\gamma}

\titleformat{\paragraph}[hang]{\normalfont\normalsize\bfseries}{\theparagraph}{1em}{}
\titlespacing*{\paragraph}{0pt}{3.25ex plus 1ex minus .2ex}{0.5em}

\newcolumntype{A}{ >{$} r <{$} @{} >{${}} l <{$} } 

\renewcommand{\thefootnote}{\fnsymbol{footnote}}

\usepackage{sepfootnotes}

\makeatletter
\def\@fnsymbol#1{%
	\ensuremath{\ifcase#1\or
	\ddagger\or			
	\mathsection\or		
	*\or		
	\dagger\or			
	\mathflat\or				
	\|\or				
	\ddagger\ddagger\or	
	\dagger\dagger\or	
	**					
	\else\@ctrerr\fi}}
\makeatother

\pgfdeclarelayer{bg0}    
\pgfdeclarelayer{bg1}
\pgfsetlayers{bg0,bg1,main}  

\begin{document}

	\begin{center}
	
		{\textbf{
        {\large The $(2+\delta)$-dimensional theory of the  electromechanics of lipid membranes:\\[4pt]  III. Constitutive models}
        }} \\
		
		\vspace{0.21in}

        \sepfootnotecontent{YO}{\href{mailto:yadomar@mit.edu}{yadomar\textit{@}mit.edu}}   
        \sepfootnotecontent{KM}{\href{mailto:kranthi@berkeley.edu}{kranthi\textit{@}berkeley.edu}}

		{\small
			Yannick A. D. Omar\,\textsuperscript{1,2,}\sepfootnote{YO},
			Zachary G. Lipel\,\textsuperscript{1,3},
			and Kranthi K. Mandadapu\,\textsuperscript{1,4,}\sepfootnote{KM} \\
		}
		\vspace{0.15in}

		\footnotesize{
			{
                $^1$
				Department of Chemical \& Biomolecular Engineering,
				University of California, Berkeley, CA 94720, USA
				\\[3pt]
                $^2$ Department of Chemical Engineering, Massachusetts Institute of Technology, Cambridge, MA 02139, USA\\[3pt]
                $^3$ Department of Chemical \& Biological Engineering, Princeton University, Princeton, NJ 08544, USA\\[3pt]
				$^4$
				Chemical Sciences Division, Lawrence Berkeley National Laboratory, CA 94720, USA
				\\
			}
		}
	\end{center}

	%
	%

	\begin{abstract}
        This article is the final part of a three-part series that develops a self-consistent theoretical framework describing the electromechanics of arbitrarily curved lipid membranes at the continuum scale. Owing to their small thickness, lipid membranes are commonly modeled as two-dimensional surfaces. However, this approach breaks down when considering their electromechanical behavior as it requires accounting for their finite thickness. To address this, we developed a new dimension reduction procedure in part 1 to derive \textit{effective} surface theories that explicitly capture the finite thickness of lipid membranes. We applied this method to dimensionally reduce Gauss' law and the electromechanical balance laws and referred to the resulting theory as $(2+\delta)$-dimensional, where $\delta$ indicates the membrane thickness. However, the $(2+\delta)$-dimensional balance laws for thin bodies derived in part 2 are general, and specific constitutive material models must be incorporated to specialize them to lipid membranes.
        In this work, we devise appropriate three-dimensional constitutive models that capture the material behavior of lipid membranes, which flow along their in-plane directions like viscous fluids but bend out-of-plane like elastic solids.
        The viscous material behavior is recovered by considering a three-dimensional Newtonian fluid model, leading to the same viscous stresses as strictly two-dimensional models of lipid membranes. The elastic resistance to bending is recovered by imposing a free energy penalty on local volume changes. While this material model does give rise to the characteristic bending resistance of lipid membranes, it differs in its higher-order curvature terms from the two-dimensional Canham-Helfrich-Evans theory. Furthermore, since lipid membranes only exhibit small mid-surface stretch, they are often considered mid-surface area incompressible. In this work, this is captured by introducing reactive stresses that give rise to an effective surface tension. Finally, we use the viscous, elastic, and reactive stresses to derive the equations of motion and boundary conditions describing the electromechanics of lipid membranes. We conclude this article by providing the equations of motion and coupling- and boundary conditions for a charged lipid membrane embedded in an electrolyte solution.
	\end{abstract}

	%
	%

	{ \hypersetup{linkcolor=black} \tableofcontents }

	%
	%
	
    \renewcommand{\thefootnote}{\arabic{footnote}}
    \setcounter{footnote}{0} 

    \section{Introduction}

This work concludes a series of three articles proposing an effective and self-consistent surface theory of the electromechanics of lipid membranes. In part 1\cite{omar2024ES}, we introduced a new dimension reduction procedure for deriving effective surface theories from their three-dimensional counterparts, and applied it to the electrostatics of lipid membranes. In part 2 \cite{omar2023BL}, we used the procedure to obtain effective surface mass, linear, and angular momentum balances for thin bodies. In this article, we specialize the effective balance laws to the viscous-elastic material response of lipid membranes while also accounting for their electromechanical coupling.\textspace

Since the early work by Canham, Helfrich, and Evans \cite{canham:1970,helfrich1973elastic,evans:1974}, the mechanics of lipid membranes are commonly formulated in terms of surface descriptions. This approach significantly simplifies solving the governing equations for lipid membranes undergoing arbitrary deformations (see e.g. \cite{brochard1975frequency,derenyi2002formation,sauer2017stabilized, gross2018hydrodynamic,torres2019modelling,sahu2020geometry,sahu2020arbitrary}). However, it cannot be trivially extended to describe the coupled electrical and mechanical behavior observed in lipid membranes \cite{angelova1988mechanism,riske2005electro,dimova2007giant,vlahovska2019electrohydrodynamics}. Specifically, surface theories cannot resolve distinct surface charges on the interfaces between lipid membranes and their surrounding fluids, as well as potential differences across and electric fields within lipid membranes. This makes surface theories unsuitable for describing Maxwell stresses within lipid membranes and in their vicinity, and consequently, for describing the electromechanics of lipid membranes.\textspace

The dimension reduction procedure proposed in part 1 \cite{omar2024ES} allows us to derive effective surface theories that capture finite thickness effects of lipid membranes by using a differential geometry framework and low-order spectral expansions. We showed that applying this method to the electrostatics of thin bodies yields a surface theory that does not suffer from the aforementioned shortcomings \cite{omar2024ES}. We refer to this theory as $(2+\delta)$-dimensional, where $\delta$ indicates the membrane thickness. To obtain a self-consistent $(2+\delta)$-dimensional theory that captures the electromechanics of lipid membranes, we applied the dimension reduction method to the mass, angular, and linear momentum balances while accounting for Maxwell stresses, and derived the corresponding equations of motion in part 2 \cite{omar2023BL}. This new electromechanical theory further showed that existing surface theories do not resolve the distinct velocities and tractions on the two interfaces between the membrane and their surrounding fluid.\textspace

The electromechanical theory derived in part 2 does not make any assumptions about the constitutive behavior of lipid membranes and is therefore generally valid for thin or shell-like bodies. Yet, strict surface theories of lipid membranes succeed at describing their in-plane viscous behavior and their elastic resistance to in-plane stretch and out-of-plane bending. So far, we have not discussed how such a viscous-elastic material response of lipid membranes can be captured by the $(2+\delta)$-dimensional theory. However, one of the advantages of the $(2+\delta)$-dimensional theory is that it directly incorporates three-dimensional constitutive models and does not require proposing two-dimensional ones. In this article, we choose appropriate three-dimensional constitutive models to specialize the electromechanical theory presented in part 2 to lipid membranes. 
Specifically, we attribute the elastic response of lipid membranes exclusively to volumetric changes, consistent with their in-plane fluidity. This model reproduces the known resistance to in-plane stretch and out-of-plane bending, albeit with a different functional form of the equations of motion than direct surface theories. Furthermore, we show that the viscous behavior and viscous-elastic coupling are captured by treating lipid membranes as three-dimensional Newtonian fluids. Additionally, in the case of mid-surface area incompressibility, we find that the dimension reduction method gives rise to an effective surface tension. Thus, we obtain an electromechanical theory that not only captures the viscous-elastic material behavior of lipid membranes similar to strict surface theories, but also accounts for their electromechanical coupling.\textspace

The remainder of this article is structured as follows. In Sec.~\ref{sec:kinematics}, we revisit the three-dimensional kinematics used to describe lipid membranes in a differential geometry framework. Subsequently, we summarize the dimension reduction method proposed in part 1 \cite{omar2024ES} and describe the dimensionally-reduced electromechanical theory derived in part 2 \cite{omar2023BL}. This prepares us to propose three-dimensional elastic and viscous constitutive models appropriate for lipid membranes in Sec.~\ref{sec:const_models}. Using these constitutive models, we derive the $(2+\delta)$-dimensional equations of motion and boundary conditions for lipid membranes in Sec.~\ref{sec:EOMs}. Section~\ref{sec:EOMElectrolyte} then presents the complete set of equations describing the electromechanics of a lipid membrane with surface charges embedded in an electrolyte. Readers interested in applying the theory derived in this series of articles may refer to the latter two sections.

\paragraph{Notation} The following notation is common among all three parts of this series of articles \cite{omar2024ES,omar2023BL}. Vectors and tensors are denoted by bold letters (e.g. $\bm{v}$) and matrices are denoted using square brackets (e.g. $\left[ B_{ij} \right]$). Subscripts and superscripts indicate covariant and contravariant components (e.g. $v_i, \, v^i$), respectively, and repeated sub- and superscripts imply Einstein's summation convention. Greek indices (e.g. $\alpha, \, \beta$) take values $\{1,2\}$ while Latin indices (e.g. $i,\,j$) take values $\{1,2,3\}$ except when used in the context of Chebyshev expansions. For some function $f$, the short-hand notations $f\leftR(\theta^\alpha\rightR)$ and $f\leftR(\theta^i\rightR)$ imply $f\leftR(\theta^1,\theta^2\rightR)$ and $f\leftR(\theta^1,\theta^2, \theta^3\rightR)$, respectively, where $\theta^1,\theta^2,\theta^3$ are curvilinear coordinates. Partial derivatives with respect to coordinates $\theta^k$ are denoted by a comma (e.g. $a_{i,k} = \partial a_i/\partial \theta^k$) and covariant derivatives of vector components $a^i$ and tensor components $A^{ij}$ in three-dimensional space are denoted by $a^i\bigr\rvert_k$ and $A^{ij}\bigr\rvert_{k}$, respectively. 
Similarly, the covariant derivatives of vector components $a^\alpha$ and tensor components $A^{\alpha\beta}$ defined on the tangent space of a surface in three dimensions are denoted by $a^\alpha_{:\gamma}$ and $A^{\alpha\beta}_{;\gamma}$, respectively. For non-symmetric, second-order tensors, it is necessary to indicate the order of indices, which is achieved here using a dot. For example, $A_{i}^{\,.\,j}$ implies that $i$ is the first index while $j$ is the second index. However, most tensors considered in this article are symmetric such that the order of co- and contravariant components is inconsequential and does not need to be indicated, i.e. $A_{i}^{\,.\,j} = A_{\,.\,i}^{j} =A_i^j\,$.

\section{Kinematics} \label{sec:kinematics}
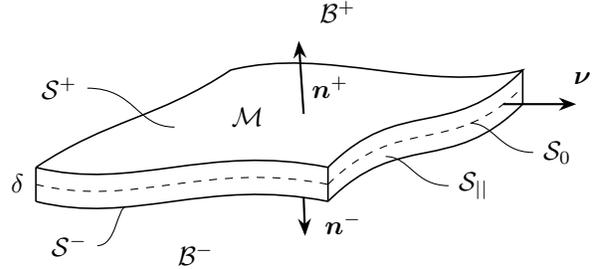
\begin{wrapfigure}{r}{0.49\textwidth}
    \centering
    \tikzset{>=latex}
\begin{tikzpicture}[font=\small, x=0.8\linewidth, y=0.8\linewidth] 

\pgfdeclarelayer{background}
\pgfdeclarelayer{foreground}
\pgfsetlayers{background,foreground}

\def \h{0.07}

\begin{pgfonlayer}{background}
\coordinate (A) at (0,0.1); 
\coordinate (B) at (0.6,0.1); 
\coordinate (C) at (1,0.3); 
\coordinate (D) at (0.4,0.3); 
\coordinate (topcenter) at (0.55, 0.28);
\coordinate (bottomcenter) at (0.55, 0.109);
\coordinate (nustart) at (0.96, 0.3);

\draw[semithick] 
      (A) .. controls +(0.2, -0.05) and +(-0.25,0.05)  .. (B)
      (B) .. controls +(0.15, 0.15) and +(-0.15, -0.15) .. (C)
      (A) -- ++(0,\h) coordinate (E)
      (B) -- ++(0,\h) coordinate (F)
      (C) -- ++(0,\h) coordinate (G)
      (D) -- ++(0,\h) coordinate (H);
      
\end{pgfonlayer}

\begin{pgfonlayer}{foreground}
\filldraw[fill=white, semithick] 
      (E) .. controls +(0.2, -0.05) and +(-0.25,0.05)  .. (F) .. controls +(0.15, 0.15) and +(-0.15, -0.15) .. (G) .. controls +(-0.2, -0.05) and +(0.2, 0.05) .. (H) .. controls +(-0.15, -0.1) and +(0.15, 0.1) .. (E);

\node [position=-7pt:7pt from A] (hnode) {$\delta$};
\node [position=-45pt:-5pt from topcenter] (ltop) {};
\node [position=18pt:17pt from B] (rbot) {};
\node [position=32pt:12.6pt from rbot] (rmid) {};
\node [position=38pt:-2pt from A] (fmid) {};

\draw[] (ltop) .. controls +(-0.1,0) and +(0.1,-0.) .. ++(-0.2,0.08) node[anchor=east] (nodeSp) {$\mathcal{S}^+$};
\draw (rbot) .. controls +(0.07,0) and +(-0.07,-0.) .. ++(0.15,-0.06) node[anchor=west] (nodeSpar) {$\mathcal{S}_{||}$};
\draw (rmid) .. controls +(0.07,0) and +(-0.07,-0.) .. ++(0.15,-0.06) node[anchor=west] (nodeS0) {$\mathcal{S}_0$}; 
\draw (fmid) .. controls +(-0.04,0) and +(0.04,0) .. +(-0.08, -0.08) node[anchor=east] (nodeSm) {$\mathcal{S}^-$}; 
\node[position=80pt:32pt from A] (nodeM) {$\mathcal{M}$};
\node[position=-70pt:35pt from C] (nodeBp) {$\mathcal{B}^+$};
\node[position=-50pt:-20pt from B] (nodeBm) {$\mathcal{B}^-$};

\draw[-{Stealth}, thick] (topcenter) -- ++(-0.01, 0.15);
\node [position=10pt:10pt from topcenter] (nplabel) {${\bm{n}^+}$};  
\draw[-{Stealth}, thick] (bottomcenter) -- ++(0.003, -0.08);
\node [position=15pt:-10pt from bottomcenter] (nmlabel) {${\bm{n}^-}$};  

\draw[-{Stealth}, thick] (nustart) -- ++(0.15, 0.);
\node [position=30pt:10pt from nustart] (nulabel) {$\bm{\nu}$};  

\draw[dashed] 
    ($(A)+(0.,\h/2)$) .. controls +(0.2, -0.05) and +(-0.25,0.05)  .. ($(B)+(0,\h/2)$)
    ($(B)+(0.,\h/2)$) .. controls +(0.15, 0.15) and +(-0.15, -0.15) .. ($(C)+(0.,\h/2)$);
\end{pgfonlayer}

\end{tikzpicture}
    \caption{Schematic of the setup used for our derivations. We consider a lipid membrane $\mathcal{M}$ with constant thickness $\delta$ separating the two bulk domains $\mathcal{B}^+$ and $\mathcal{B}^-$. Its top, bottom, and lateral bounding surfaces are denoted by $\mathcal{S}^+$, $\mathcal{S}^-$, and $\mathcal{S}_{||}$, and their outward pointing normals are $\bmn^+$, $\bmn^-$, and $\bm{\nu}$, respectively. Similarly, the mid-surface is denoted by $\mathcal{S}_0$ with normal vector $\bmn$.}
    \label{fig:schematic_setup}
\end{wrapfigure}
In this section, we briefly summarize the kinematics we used to describe arbitrarily curved lipid membranes in parts 1 and 2 \cite{omar2024ES,omar2023BL}. To this end, consider a lipid membrane $\mathcal{M}$ with constant thickness $\delta$ that is embedded into a bulk domain (Fig.~\ref{fig:schematic_setup}). The segments of the bulk domain that lie above and below $\mathcal{M}$ are denoted by $\mathcal{B}^+$ and $\mathcal{B}^-$, respectively. The surfaces $\mathcal{S}^\pm$ form the interfaces between $\mathcal{M}$ and $\mathcal{B}^\pm$ and have normal vectors $\bm{n}^\pm$ pointing from $\mathcal{M}$ towards $\mathcal{B}^\pm$ as shown in Fig.~\ref{fig:schematic_setup}. Furthermore, $\mathcal{S}_0$ denotes the mid-surface between the interfaces $\mathcal{S}^\pm$ with normal vector $\bmn$, and $\mathcal{S}_{||}$ is the lateral bounding surface with outward pointing normal $\bm{\nu}$.\textspace

Arbitrarily curved and deforming lipid membranes are conveniently described using a differential geometry framework \cite{steigmann1999fluid,aris2012vectors,rangamani2013interaction,sahu2017irreversible,itskov2019tensor}. To this end, we introduce time-dependent, curvilinear coordinates $(\theta^{1}, \theta^2) \in \Omega$ parametrizing the mid-surface $\mathcal{S}_0$ and $\theta^3 \in \Xi$ parametrizing $\mathcal{M}$ along the thickness, where $\Omega$ and $\Xi$ denote appropriate parametric spaces. This allows us to express $\mathcal{M}$ using the position vector $\bmx\leftR(\theta^i,t\rightR)$ as
\begin{align}
    \mathcal{M} = \left\{ \bmx\leftR(\theta^i,t\rightR): (\theta^{1}, \theta^2, \theta^3) \in \Omega \times \Xi \right\}~.
\end{align}
Similarly, the mid-surface $\mathcal{S}_0$ can be described in terms of the parametrization as 
\begin{align}
    \mathcal{S}_0 = \left\{ \bmx\leftR(\theta^i,t\rightR): (\theta^{1}, \theta^2, 0) \in \Omega \times \Xi \right\} = \left\{ \bmx_0\leftR(\theta^\alpha,t\rightR): (\theta^{1}, \theta^2) \in \Omega \right\}~,
\end{align}
where we introduced the mid-surface position vector $\bmx_0\leftR(\theta^\Ga,t\rightR) \in \mathcal{S}_0$. We now assume that the body $\mathcal{M}$ can be described by expressing the position vector as
\begin{align}
    \bmx\leftR(\theta^i,t\rightR) = \bmx_0\leftR(\theta^\Ga,t\rightR) + \bmn\leftR(\theta^\Ga,t\rightR) \theta^3~, \label{eq:xform}
\end{align}
which is discussed in extensive detail in part 2 \cite{omar2023BL}. Equation~\eqref{eq:xform} together with the assumption of constant thickness implies $\theta^3 \in (-\delta/2,\delta/2)$ and that $\theta^3$ is independent of time.\textspace

To describe the kinematics of $\mathcal{M}$ in more detail, we introduce a number of additional geometric quantities. The parametrization of the mid-surface induces the tangent vectors $\bma_\Ga$ to the mid-surface, 
\begin{align}
    \bma_\Ga = \frac{\partial \bmx_0}{\partial \theta^\Ga}\biggr\rvert_t~, \label{eq:bma_def}
\end{align}
which facilitates evaluation of the normal vector to the mid-surface,
\begin{align}
    \bm{n} = \frac{\bma_1 \times \bma_2}{||\bma_1 \times \bma_2||}~.\label{eq:bmn_def}
\end{align}
The tangent and normal vectors in Eqs.~\eqref{eq:bma_def} and~\eqref{eq:bmn_def} form a basis of $\mathbb{R}^3$ that we denote by $\left\{\check{\bmg}_i\right\}_{i=1,2,3} = \left\{\bma_1,\bma_2, \bmn\right\}$.
Equation~\eqref{eq:bma_def} further induces the covariant and contravariant metric tensors,
\begin{align}
    a_{\Ga\Gb} &= \bma_\Ga \cdot \bma_\Gb~, \\
    \left[ a^{\Ga\Gb} \right] &= \left[ a_{\Ga\Gb} \right]^{-1}~,
\end{align}
which let us alter between covariant and contravariant components of purely tangential vectors and tensors. Accordingly, we can define the contravariant in-plane basis vectors as
\begin{align}
    \bma^\Ga = a^{\Ga\Gb} \bma_\Gb~. 
\end{align}
The covariant components of the mid-surface curvature tensor $\bmb$ are given by
\begin{align}
    b_{\Ga\Gb} = \bma_{\Ga,\Gb} \cdot \bmn~,
\end{align}
and the two invariants of $\bmb$ are the mean curvature $H$ and Gaussian curvature $K$ defined as 
\begin{align}
    H = \frac{1}{2} \tr{\bmb} &= \frac{1}{2} b^\Ga_\Ga~, \label{eq:Hdef}\\
    K = \det{\bmb} &= \det{\left[b^\Ga_\Gb\right]}~. \label{eq:Kdef}
\end{align}

Equations~\eqref{eq:bma_def}--\eqref{eq:Kdef} only characterize the mid-surface geometry of $\mathcal{M}$. However, we can also find the basis vectors away from the mid-surface, defined as $\bmg_i = \frac{\partial \bmx}{\partial \theta^i}\bigr\rvert_t$. According to  Eq.~\eqref{eq:xform}, these are given by
\begin{align}
    \bm{g}_\alpha &= \bm{a}_\alpha - b_{\alpha}^{\beta} \bm{a}_\beta \theta^3~, \label{eq:g_a}\\
    \bm{g}_3 &= \bm{n}~, \label{eq:g_3}
\end{align}
where we applied the Weingarten formula $\bm{n}_{,\alpha} = -b_\alpha^\beta \bm{a}_\beta$. Thus, the three dimensional metric tensor takes the form 
\begin{align}
    g_{ij} = 
    \begin{bmatrix}
    \left[g_{\Ga\Gb}\right] & \bm{0} \\
    \bm{0} & 1
    \end{bmatrix}_{ij}~. \label{eq:gij_diag}
\end{align}
From Eqs.~\eqref{eq:g_a} and~\eqref{eq:gij_diag}, it is apparent that the componets $g^{\Ga\Gb}$ are rational polynomials in $\theta^3$, implying that the contravariant basis vector $\bm{g}^\Ga$ are also rational polynomials in $\theta^3$. For the purpose of applying the dimension reduction procedure introduced in part 1 \cite{omar2024ES}, it is convenient to express $\bmg^\Ga$ as a series expansion (see Supplementary Material of part 2 \cite{omar2023BL}),
\begin{align}
    \bmg^\Ga = \sum_{m=0}^\infty \bm{a}^\alpha \cdot {\bmb}^{m}  \left(\theta^3\right)^m~. \label{eq:ga_full_expansion}
\end{align}
Furthermore, the block-diagonal structure of $g_{ij}$ in Eq.~\eqref{eq:gij_diag} implies that the third basis vector is invariant to coordinate transformations, i.e.
\begin{align}
    \bm{g}^3 = \bm{g}_3 = \bm{n}~. 
\end{align}
Using the definition of the basis vectors, we can define the in-plane and three-dimensional identity tensors, respectively, as
\begin{align}
    \bmi &= \bma_\Ga \otimes \bma^\Ga~, \\
    \bm{1} &= \bmg_\Ga \otimes \bmg^\Ga~.
\end{align}

The geometric quantities introduced above now allow us to describe the choice of permissible deformations and velocities. Specifically, we restrict the deformations to satisfy Kirchhoff-Love kinematics: Any material point along the normal to the mid-surface remains on this normal and maintains a constant distance from the mid-surface during deformation. This choice implies that the velocity takes the form
\begin{align}
    {\bm{v}}\leftR(\theta^i,t\rightR)  &= 
    {\bm{v}}_0\leftR(\theta^\alpha,t\rightR) + {\bm{v}}_1\leftR(\theta^\alpha,t\rightR)\, \theta^3 \label{eq:velo_eulerian_vec}\\
    &= v_0^\alpha \bm{a}_\alpha + v_0^3 \bm{n} + v_1^\alpha \bm{a}_\alpha \, \theta^3~, \label{eq:velo_eulerian}
\end{align}
where
\begin{align}
    \bmv_1 &= \dot{\bmn}~, \\
    v_1^\alpha &= \dot{\bmn} \cdot \bma^\Ga = -\left(v^{3}_{0,\Gb} + v_0^{\lambda} b_{\lambda\Gb}\right) a^{\Gb\Ga}~, \label{eq:v1_Eulerian}
\end{align}
which is a result discussed in detail in part 2 \cite{omar2023BL}. Here, $v_0^\Ga\leftR(\theta^\beta,t\rightR)$ and $v_0^3\leftR(\theta^\beta,t\rightR)$ denote the in-plane and normal mid-surface velocity components, respectively, and are the unknowns of our theory.\textspace

\section{Summary of the $(2+\delta)$-dimensional Balance Laws} \label{sec:2pd_summary}
In the following, we provide a brief summary of the $(2+\delta)$-dimensional electrostatics equations derived in part 1 \cite{omar2024ES}, and the $(2+\delta)$-dimensional mass balance and equations of motion derived in part 2 \cite{omar2023BL}. We begin by recalling that a sufficiently well-behaved function $f\leftR(\theta^i,t\rightR)$ can be expanded along the thickness direction $\theta^3$ in terms of Chebyshev polynomials $\ChebT{k}{\Theta}$, $\Theta = \theta^3/ (\delta/2) \in (-1,1)$, as
\begin{align}
    f\leftR(\theta^i,t \rightR) = \sum_{k=0}^\infty f_k \leftR(\theta^\Ga, t\rightR) \ChebT{k}{\Theta}~, 
\end{align}
allowing us to separate the tangential and normal directions of the mid-surface. This expansion can be similarly repeated for vector- and tensor-valued functions. In the derivation of the $(2+\delta)$-dimensional theory, we also used that Chebyshev polynomials are orthogonal, that is 
\begin{align}
	\langle \ChebT{i}{\Theta}, \ChebT{j}{\Theta} \rangle = \frac{1}{\pi} \int_{-1}^1 \ChebT{i}{\Theta}\ChebT{j}{\Theta} \, \frac{1}{\sqrt{1-\Theta^2}} \diff{\Theta} = \beta_i \delta_{ij}~, \label{eq:InnerProd_def}
\end{align}
with
\begin{align}
    \beta_i = 
    \begin{cases}
    1 \quad &\text{if } i = 0~, \\
    \frac{1}{2} \quad &\text{otherwise}~.
    \end{cases}
\end{align}
Using these preliminaries, we next state Gauss' law and the mass, linear momentum, and angular momentum balances and their $(2+\delta)$-dimensional forms. The interested reader is referred to parts 1 and 2 for detailed derivations of these results \cite{omar2024ES, omar2023BL}. 

\paragraph{The $(2+\delta)$-dimensional Gauss' law}
In part 1 \cite{omar2024ES}, we started from the three-dimensional Gauss' law in the membrane
\begin{alignat}{4}
     \varepsilon_\mathcal{M}\Delta {\phi}_\mathcal{M} &= 0~, \label{eq:gauss_M} 
\end{alignat}
and the electrostatic jump conditions on the top and bottom surfaces $\mathcal{S}^\pm$
\begin{align}
    \bm{n}^\pm \cdot \llbracket \varepsilon {\bm{e}} \rrbracket^\pm &= \sigma^\pm~, \label{eq:jumpSp}
\end{align}
to obtain the $(2+\delta)$-dimensional electrostatic theory. In Eq.~\eqref{eq:gauss_M}, $\varepsilon_\mathcal{M}$ denotes the permittivity of the membrane, $\Delta$ is the three-dimensional Laplacian and ${\phi}_\mathcal{M}$ is the electric potential in the membrane. In Eq.~\eqref{eq:jumpSp}, $\sigma^\pm$ denotes the surface charge density on the top and bottom surfaces and the jumps $\llbracket \varepsilon {\bm{e}} \rrbracket^\pm$ are defined as
\begin{align}
    \llbracket \varepsilon {\bm{e}} \rrbracket^\pm &= \varepsilon_{\mathcal{B}^\pm} \bme_{\mathcal{B}^\pm}\rvert_{\mathcal{S}^\pm} - \varepsilon_{\mathcal{M}} \bme_{\mathcal{M}}\rvert_{\mathcal{S}^\pm}~, \label{eq:ES_jump}
\end{align}
with $\varepsilon_{\mathcal{B}^\pm}$ and $\bme_{\mathcal{B}^\pm}$ denoting the bulk permittivities and electric fields, respectively, and $\bme_\mathcal{M} = -\grad{\phi_\mathcal{M}}$. To close these equations with appropriate boundary conditions on the lateral surface $\mathcal{S}_{||}$ (see Fig.~\ref{fig:schematic_setup}), we divide $\mathcal{S}_{||}$ into parts with prescribed potential ($\mathcal{S}_{||\mathrm{D}}^\mathrm{ES}$) and electric field ($\mathcal{S}_{||\mathrm{N}}^\mathrm{ES}$), i.e.
\begin{align}
    \mathcal{S}_{||}^\mathrm{ES} &= \mathcal{S}_{||\mathrm{D}}^\mathrm{ES} \cup \mathcal{S}_{||\mathrm{N}}^\mathrm{ES}~, \label{eq:Spar_DES} \\
    \emptyset &= \mathcal{S}_{||\mathrm{D}}^\mathrm{ES} \cap \mathcal{S}_{||\mathrm{N}}^\mathrm{ES}~.\label{eq:Spar_NES}
\end{align}
This allows us to prescribe the boundary conditions
\begin{alignat}{2}
    {\phi}_{\mathcal{M}} &= \bar{\phi}_\mathcal{M}~&&,\quad \forall {\bm{x}} \in \mathcal{S}_{||\mathrm{D}}^\mathrm{ES}~, \label{eq:DBCES_Spar}\\
    -\bm{\nu} \cdot \grad{{\phi}_\mathcal{M}} &= \bar{e}~&&,\quad \forall {\bm{x}} \in \mathcal{S}_{||\mathrm{N}}^\mathrm{ES}~, \label{eq:NBCES_Spar}
\end{alignat}
where $\bar{\phi}_\mathcal{M}$ and $\bar{e}$ indicate the prescribed potential and normal component of the electric field, respectively. Furthermore, in writing Eqs.~\eqref{eq:DBCES_Spar} and~\eqref{eq:NBCES_Spar}, we assume that the same type of boundary condition, i.e. either the electric potential or field, is chosen throughout the thickness at any point of the boundary.\textspace

By expanding the potential in the membrane to second order,
\begin{align}
    \phi_{\mathcal{M}} &= \mathlarger{\sum}_{k = 0}^{2} ~ \phik{k} \ChebT{k}{\Theta}~, \label{eq:chebphi}
\end{align}
we obtain the $(2+\delta)$-dimensional form of Gauss' law as 
\begin{align}
    \varepsilon_\mathcal{M} \laplaces{\phi_0} - 4 C_\mathcal{M} \phi_1 H +  \frac{16}{\delta} C_\mathcal{M} \phi_2 = 0~, \label{eq:surface_gauss} 
\end{align}
with the surface Laplacian $\Delta_\mathrm{s}\leftR(\bullet\rightR) =  \left( \left( \bullet \right)_{,\alpha}\right)_{:\beta} a^{\Ga\Gb}$ and the effective specific membrane capacitance per unit area $C_\mathcal{M} = \varepsilon_\mathcal{M}/\delta$. Equation~\eqref{eq:surface_gauss} is the governing equation for $\phi_0$, wherein the remaining expansion coefficients $\phi_1$ and $\phi_2$ are obtained from the jump conditions in Eqs.~\eqref{eq:ES_jump} as 
\begin{align}
    \phi_1 &= -\frac{1}{2 C_\mathcal{M}} \left(\bm{n} \cdot \langle \varepsilon_\mathcal{B} \bm{e}_{\mathcal{B}} \rangle^\mathcal{M} - \frac{1}{2}\left( \sigma^+ - \sigma^- \right)\right)~, \label{eq:phi1} \\[6pt]
    \phi_2 &=  -\frac{1}{16 C_\mathcal{M}} \left(\bm{n} \cdot \llbracket \varepsilon_\mathcal{B}\bm{e}_{\mathcal{B}} \rrbracket^\mathcal{M} - \left(\sigma^+ + \sigma^-\right)\right)~, \label{eq:phi2}
\end{align}
with the average $\langle \varepsilon_{\mathcal{B}} \bm{e}_{\mathcal{B}} \rangle^\mathcal{M} = \frac{1}{2}\left(\varepsilon_{\mathcal{B}^+}\bm{e}_{\mathcal{B}^+}\rvert_{\mathcal{S}^+} + \varepsilon_{\mathcal{B}^-}\bm{e}_{\mathcal{B}^-}\rvert_{\mathcal{S}^-} \right)$ and jump $\llbracket \varepsilon_{\mathcal{B}}\bm{e}_{\mathcal{B}} \rrbracket^\mathcal{M} = \varepsilon_{\mathcal{B}^+}\bm{e}_{\mathcal{B}^+}\rvert_{\mathcal{S}^+} - \varepsilon_{\mathcal{B}^-}\bm{e}_{\mathcal{B}^-}\rvert_{\mathcal{S}^-}$. 
Furthermore, the boundary conditions of the $(2+\delta)$-dimensional theory corresponding to Eqs.~\eqref{eq:DBCES_Spar} and~\eqref{eq:NBCES_Spar} are
\begin{alignat}{2}
    \phi_0 &= \bar{\phi}_{\mathcal{M}0}~,&&\quad \forall {\bm{x}}_0 \in \partial \mathcal{S}_{0\mathrm{D}}^\mathrm{ES}~, \label{eq:DBCES_Spar_red} \\
    -{\nu}^\mu \left( \phi_{0,\mu} + \frac{\delta}{4} \phi_{1,\beta} b^\beta_\mu + \frac{\delta^2}{16} \phi_{2,\beta} b^\beta_\gamma b^\gamma_\mu \right) &= \bar{e}_0~,&&\quad \forall {\bm{x}}_0 \in \partial \mathcal{S}_{0\mathrm{N}}^\mathrm{ES}~, \label{eq:NBCES_Spar_red}
\end{alignat}
where $\bar{\phi}_{\mathcal{M}0} = \langle \bar{\phi}_{\mathcal{M}}, \ChebT{0}{\Theta} \rangle$ and $\bar{e}_0 = \langle \bar{e}, \ChebT{0}{\Theta} \rangle$ are the zeroth-order components of the prescribed electric potential and field, respectively, and $\partial \mathcal{S}_{0\mathrm{D}}^\mathrm{ES} = \mathcal{S}_{||\mathrm{D}}^\mathrm{ES} \cap \partial \mathcal{S}_0$ and $\partial \mathcal{S}_{0\mathrm{N}}^\mathrm{ES} = \mathcal{S}_{||\mathrm{N}}^\mathrm{ES} \cap \partial \mathcal{S}_0$ with $\partial \mathcal{S}_0$ being the boundary of $\mathcal{S}_0$.

\paragraph{The $(2+\delta)$-dimensional mass balance}
From the three-dimensional mass balance, 
\begin{align}
    \frac{\ddiff{\rho}}{\ddiff t} + \rho \divv{\bmv} = 0~,
\end{align}
it is apparent that we require expansions of the density $\rho$ and velocity $\bmv$ in terms of Chebyshev polynomials. Here, we use a series expansion for the density
\begin{align}
    \rho\leftR(\theta^i,t \rightR) &= \sum_{k= 0}^\infty \rho_k\leftR(\theta^\Ga, t\rightR) \ChebT{k}{\Theta}~, \label{eq:density_expansion_eulerian}
\end{align}
and the first-order expansion in Eq.~\eqref{eq:velo_eulerian_vec} for the velocity, consistent with Kirchhoff-Love kinematics,
\begin{align}
    \bmv\leftR(\theta^i,t\rightR) &= \bmv_0\leftR(\theta^\Ga,t\rightR)\ChebT{0}{\Theta} + \frac{\delta}{2}\bmv_1\leftR(\theta^\Ga,t\rightR)\ChebT{1}{\Theta}~.  \label{eq:velo_chebT_expansion}
\end{align}
With these definitions, the $(2+\delta)$-dimensional mass balance takes the well-known form found in surface theories of lipid membranes \cite{scriven1960dynamics, sahu2017irreversible},
\begin{align}
    \frac{\ddiff \rho_\mathrm{s} }{\ddiff t} +  \rho_\mathrm{s} \left( v_{0:\Ga}^\Ga- 2  v_0^3 H \right)= 0~, \label{eq:2pd_mass_balance}
\end{align}
where $\rho_\mathrm{s} = \delta \rho_0$ and $v_{0:\Ga}^\Ga$ denotes the covariant derivative of $ v_{0}^\Ga$.

\paragraph{The $(2+\delta)$-dimensional linear and angular momentum balances}
To derive the $(2+\delta)$-dimensional equations of motion, we begin from the three-dimensional linear and angular momentum balances for non-polar media,
\begin{align}
    \rho \frac{\ddiff \bmv}{\ddiff t} &= \divv{\bm{\sigma}^T} + \bmf~, \label{eq:3Dlinmom} \\
    \bmg_i \times \bmT^i &= \bm{0}~, \label{eq:3Dangmom}
\end{align}
where $\bmf$ is the body force per unit volume, $\bm{\sigma}$ is the purely mechanical Cauchy stress and 
\begin{align}
    \bmT^i = \bm{\sigma}^T\bmg^i~, \label{eq:stressvec_def}
\end{align}
are the corresponding stress vectors\footnote{In the theory of thin bodies, it is often convenient to work with stress vectors rather than stress tensors.} \cite{gurtin2010mechanics}. We note that Eqs.~\eqref{eq:3Dlinmom}--\eqref{eq:stressvec_def} are devoid of electromechanical effects as will be discussed in more detail next.\textspace

To that end, we introduce the Maxwell stresses, which induce electromechanical effects, under the assumption of electrostatics and linear dielectric materials for the membrane and bulk~\cite{melcher1981continuum,kovetz2000electromagnetic,edmiston2011analysis,fong2020transport}, respectively,
\begin{align}
    \bm{\sigma}_\mathrm{M} &= \varepsilon_\mathcal{M}\left(\bme \otimes \bme - \frac{1}{2}\left(\bme \cdot \bme\right) \bm{1}\right)~, \\
    \bm{\sigma}_{\mathrm{M}\mathcal{B}^\pm} &= \varepsilon_{\mathcal{B}^\pm}\left(\bme_{\mathcal{B}^\pm} \otimes \bme_{\mathcal{B}^\pm} - \frac{1}{2}\left(\bme_{\mathcal{B}^\pm} \cdot \bme_{\mathcal{B}^\pm}\right) \bm{1}\right)~.
\end{align}
\begin{wrapfigure}{r}{0.55\textwidth}
    \centering
    \begin{annotationimage}[]{width=0.9\linewidth}{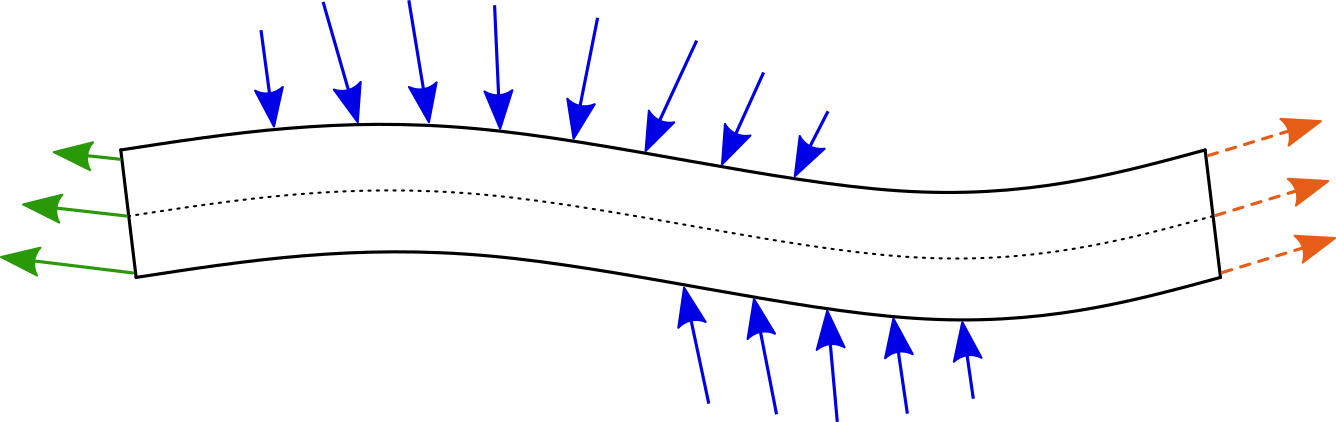}
        \imagelabelset{
                coordinate label style/.style = {
                rectangle,
                fill = none,
                text = black,
                font = \normalfont
        }}
        \draw[coordinate label = {$\color{myblue} \bm{t}^+$ at (0.6,0.92)}];
        \draw[coordinate label = {$\color{myblue} \bm{t}^-$ at (0.47,0.15)}];
        \draw[coordinate label = {$\color{mygreen} \bar{\bm{t}}$ at (-0.01,0.55)}];
        \draw[coordinate label = {$\color{myorange} \bm{\nu} $ at (0.95,0.3)}];
    \end{annotationimage}
    \caption{Mechanical tractions acting on the top ($\bmt^+$) and bottom surfaces ($\bmt^-$), and prescribed total tractions acting on the lateral bounding surface ($\bar{\bmt}$) with outward pointing normal $\bm{\nu}$.}
    \label{fig:BC_schematic}
\end{wrapfigure}
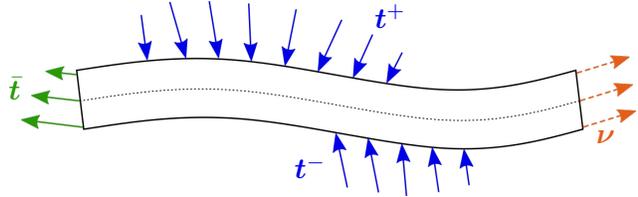
In addition, we define the total stress tensors in the membrane and bulk, respectively, as
\begin{align}
    \bbar{\bm{\sigma}} &= \bm{\sigma} + \bm{\sigma}_\mathrm{M}~, \label{eq:barbar_sigma} \\
    \bbar{\bm{\sigma}}_{\mathcal{B}^\pm} &= \bm{\sigma}_{\mathcal{B}^\pm} + \bm{\sigma}_{\mathrm{M}\mathcal{B}^\pm}~. \label{eq:barbar_sigmaB}
\end{align}

In general\footnote{If electrostatics were not assumed, additional electromechanical terms would enter the linear momentum balance~\cite{kovetz2000electromagnetic,steigmann2009formulation,fong2020transport}.}, it is the total stress $\bbar{\bm{\sigma}}$ that enters in the linear and angular momentum balances in Eqs.~\eqref{eq:3Dlinmom} and \eqref{eq:3Dangmom}~\cite{kovetz2000electromagnetic,steigmann2009formulation,fong2020transport}. However, the Maxwell stresses are symmetric and therefore vanish from the angular momentum balance in Eq.~\eqref{eq:3Dangmom}. Furthermore, the interior of lipid membranes is devoid of free charges, rendering the Maxwell stresses in the membrane divergence-free \cite{kovetz2000electromagnetic}, i.e.
\begin{align}
    \divv{\bm{\sigma}_\mathrm{M}} &= 0~.
\end{align}
Thus, the Maxwell stress $\bm{\sigma}_\mathrm{M}$ does not enter Eq.~\eqref{eq:3Dlinmom} either.
In contrast, however, Maxwell stresses enter the traction boundary conditions, and care must be taken to ensure an appropriate balance of forces at the membrane-bulk interfaces.\textspace 

With the notation of Eqs.~\eqref{eq:barbar_sigma} and~\eqref{eq:barbar_sigmaB}, we can state the traction boundary conditions on the top and bottom bounding surfaces $\mathcal{S}^\pm$ as\footnote{We do not discuss the no-slip/no-penetration conditions on $\mathcal{S}^\pm$ here as we assume they are enforced using the governing equations of the bulk. This is discussed in more detail in part 2 \cite{omar2023BL}.}
\begin{alignat}{2}
    \bbar{\bm{\sigma}}^T\bigr\rvert_{\mathcal{S^\pm}}{\bm{n}}^\pm &= \bbar{\bm{\sigma}}_{\mathcal{B}^\pm}^T\bigr\rvert_{\mathcal{S^\pm}}{\bm{n}}^\pm~, \quad &&\forall \bm{x} \in \mathcal{S}^\pm~, \label{eq:tract_cont_0}
\end{alignat}
or, equivalently, 
\begin{align}
    \bm{\sigma}^T\bigr\rvert_{\mathcal{S^\pm}}{\bm{n}}^\pm + \bm{\sigma}_{\mathrm{M}}^T\bigr\rvert_{\mathcal{S^\pm}}{\bm{n}}^\pm = \bm{t}^\pm + \bm{\sigma}_{\mathrm{M}\mathcal{B}^\pm}^T\bigr\rvert_{\mathcal{S^\pm}}{\bm{n}}^\pm~, \quad \forall \bm{x} \in \mathcal{S}^\pm~, \label{eq:tract_cont_1}
\end{align}
with $\bmt^\pm$ denoting the mechanical contributions to the tractions acting on $\mathcal{S}^\pm$ from the bulk fluid, as shown in Fig.~\ref{fig:BC_schematic}. We also introduce boundary conditions on the lateral surface $\mathcal{S}_{||}$. To this end, we split $\mathcal{S}_{||}$ into two non-overlapping domains $\mathcal{S}_{||\mathrm{D}}^\mathrm{EM}$ and $\mathcal{S}_{||\mathrm{N}}^\mathrm{EM}$ where we impose the velocity $\bar{\bmv}$ and tractions $\bar{\bmt}$, respectively,
\begin{alignat}{2}
    \bm{v} &= \bar{\bm{v}}~, \quad &&\forall \bm{x} \in \mathcal{S}_{||\mathrm{D}}^\mathrm{EM}~, \label{eq:Spar_DEM}\\
    \bbar{\bm{\sigma}}^T \bm{\nu} &= \bar{\bm{t}}~, \quad &&\forall \bm{x} \in \mathcal{S}_{||\mathrm{N}}^\mathrm{EM}~,\label{eq:Spar_NEM}
\end{alignat}
analogous to the electric field and potential boundary conditions in Eqs.~\eqref{eq:DBCES_Spar} and~\eqref{eq:NBCES_Spar}.\textspace

To derive the dimensionally-reduced equations of motion, we use series expansions for the stress tensor and body force in the membrane,
\begin{align}
    \bm{\sigma}\leftR(\theta^i,t\rightR) &=\check{\sigma}^{ij} \check{\bm{g}}_i\leftR(\theta^\alpha,t\rightR) \otimes \check{\bm{g}}_j\leftR(\theta^\alpha,t\rightR) \label{eq:sig_check_expansion}\\
    &=\sum_{k=0}^{\infty} \check{\sigma}_k^{ij}\leftR(\theta^\alpha,t\rightR) \, \check{\bm{g}}_i\leftR(\theta^\alpha,t\rightR) \otimes \check{\bm{g}}_j\leftR(\theta^\alpha,t\rightR) \ChebT{k}{\Theta} \label{eq:stress_tensor_expansion_bar}~, \\
    \bmf\leftR(\theta^i,t\rightR) &= \sum_{k=0}^{\infty} {\bm{f}}_k\leftR(\theta^\alpha,t\rightR) \ChebT{k}{\Theta}~. \label{eq:expansion_bodyf}
\end{align}
From Eqs.~\eqref{eq:ga_full_expansion}, \eqref{eq:stressvec_def} and~\eqref{eq:stress_tensor_expansion_bar}, it then follows that the stress vectors can be expanded as 
\begin{align}
    \bmT^i = \sum_{k=0}^\infty \bmT_k^i\leftR(\theta^\Ga,t\rightR) \ChebT{k}{\Theta}~,
\end{align}
with the expansion coefficients given by (see part 2 \cite{omar2023BL} for details)
\begin{align}
    \bm{T}^\alpha_n &=  \check{\bm{g}}_\beta \cdot \frac{1}{2} \left( 
    \sum_{k = 0}^n \check{\sigma}_{n-k}^{\beta j}  \langle  \bm{g}^\alpha, \ChebT{k}{\Theta} \rangle + 
    \sum_{k=0}^\infty  \check{\sigma}_{n+k}^{\beta j}  \langle \bm{g}^\alpha, \ChebT{k}{\Theta} \rangle +
    \sum_{\substack{n> 0 \\ k = n}}^\infty  \check{\sigma}_{k-n}^{\beta j}  \langle  \bm{g}^\alpha, \ChebT{k}{\Theta} \rangle \right) 
    ~ \check{\bm{g}}_j ~, \label{eq:Talpha_n_general}\\[6pt]
    \bm{T}_n^3 &= \check{\sigma}^{3j}_n \check{\bm{g}}_j~. \label{eq:T3_n_general}
\end{align}
Expressions for the inner product $\langle  \bm{g}^\alpha, \ChebT{k}{\Theta} \rangle$ were derived in part 2 \cite{omar2023BL} and are summarized in Sec.~1 of the SM.\textspace

The $(2+\delta)$-dimensional linear momentum balance can now be written in terms of the stress vector expansions. To this end, we define the in-plane stresses $N^{\Ga\Gb}$,
\begin{align}
    N^{\Ga\Gb} &= \delta \langle \bmT^\Ga \cdot \bma^\Gb, \ChebTR{0}\rangle = \delta \bmT_0^\Ga \cdot \bma^\Gb~, \label{eq:Nab_def}
\end{align}
in-plane moments $M^{\Ga\Gb}$, 
\begin{align}
    M^{\Ga\Gb} &= -\frac{\delta^2}{2} \langle \bmT^\Ga \cdot \bma^\Gb, \ChebTR{1}\rangle = -\frac{\delta^2}{4} \bmT_1^\Ga \cdot \bma^\Gb ~, \label{eq:Mab_def}
\end{align}
and transversal shear stresses $S^\Ga$,
\begin{align}
    S^\Ga &= \delta \langle \bmT^\Ga \cdot \bmn, \ChebTR{0}\rangle = \delta \bmT_0^\alpha \cdot \bmn ~. \label{eq:Sa_def}
\end{align}
In addition, it is convenient to use the short-hand notation $\bmf_\mathrm{s} = \delta \bmf_0$ for the surface body force per unit area, where $\bmf_0$ is the zeroth-order coefficient of the body force per unit volume. 
Using these definitions, we obtain the $(2+\delta)$-dimensional form of the linear momentum balance, with the in-plane components given as
\begin{align}
    \rho_\mathrm{s}\dot{\bmv}_0 \cdot \bma^\Ga &= N^{\gamma \alpha}_{;\gamma} - S^\gamma b_\gamma^\alpha + 2H_{,\Gg} M^{\Gg\Ga}
    + \left(\bm{t}^+ + \bm{t}^-\right)\cdot \bm{a}^\Ga \nonumber \\ 
    &\hspace{1cm} + \bm{a}^\Ga \cdot \left\llbracket \left\llbracket \bm{\sigma}^T_\mathrm{M} \right\rrbracket \right\rrbracket^\mathcal{M} \bm{n} 
    - 2\delta H \left( \frac{1}{2}\left(\bm{t}^+ - \bm{t}^-\right) + \left\langle \left\llbracket \bm{\sigma}^T_\mathrm{M} \right\rrbracket \right\rangle^\mathcal{M} \bm{n}\right)\cdot\bm{a}^\Ga + \bm{f}_\mathrm{s} \cdot \bm{a}^\Ga~, \label{eq:2pd_linmom_inplane}
\end{align}
and the normal component as
\begin{align}
    \rho_\mathrm{s}\dot{\bmv}_0 \cdot \bmn &= N^{\alpha\beta}b_{\alpha\beta} + S^\alpha_{:\alpha}
    + \left(\bm{t}^+ + \bm{t}^-\right)\cdot \bm{n}  \nonumber \\
    & \hspace{1cm} + \bm{n} \cdot \left\llbracket \left\llbracket \bm{\sigma}^T_\mathrm{M} \right\rrbracket \right\rrbracket^\mathcal{M} \bm{n}
    - 2\delta H \left(\frac{1}{2}\left(\bm{t}^+ - \bm{t}^-\right) + \left\langle \left\llbracket \bm{\sigma}^T_\mathrm{M} \right\rrbracket \right\rangle^\mathcal{M} \bm{n}\right)\cdot\bm{n} + \bm{f}_{\mathrm{s}}\cdot \bmn~.\label{eq:2pd_linmom_shape}
\end{align}
In Eqs.~\eqref{eq:2pd_linmom_inplane} and~\eqref{eq:2pd_linmom_shape}, we introduced the jumps and averages of the Maxwell stress jumps,
\begin{align}
    \left\langle \left\llbracket \bm{\sigma}^T_\mathrm{M} \right\rrbracket \right\rangle^\mathcal{M} &= \frac{1}{2}\left( \left( \bm{\sigma}_{\mathrm{M}\mathcal{B}^+}\bigr\rvert_{\mathcal{S}^+} - \bm{\sigma}_{\mathrm{M}}\bigr\rvert_{\mathcal{S}^+}\right) + \left(\bm{\sigma}_{\mathrm{M}\mathcal{B}^-}\bigr\rvert_{\mathcal{S}^-} - \bm{\sigma}_{\mathrm{M}}\bigr\rvert_{\mathcal{S}^-}\right) \right)~, \label{eq:sigM_avg}\\
    \left\llbracket \left\llbracket \bm{\sigma}^T_\mathrm{M} \right\rrbracket \right\rrbracket^\mathcal{M} &= \left( \bm{\sigma}_{\mathrm{M}\mathcal{B}^+}\bigr\rvert_{\mathcal{S}^+} - \bm{\sigma}_{\mathrm{M}}\bigr\rvert_{\mathcal{S}^+}\right) - \left(\bm{\sigma}_{\mathrm{M}\mathcal{B}^-}\bigr\rvert_{\mathcal{S}^-} - \bm{\sigma}_{\mathrm{M}}\bigr\rvert_{\mathcal{S}^-}\right)~. \label{eq:sigM_diff}
\end{align}
Furthermore, the inertial terms on the left-hand sides of Eqs.~\eqref{eq:2pd_linmom_inplane} and~\eqref{eq:2pd_linmom_shape} can be expressed in terms of the velocity components as 
\begin{align}
    \rho_\mathrm{s}\dot{\bmv}_0 \cdot \bma^\Ga &= \rho_\mathrm{s}\left( v^\Ga_{0,t} +  v_0^\Gb v_{0:\Gb}^\Ga - 2v_0^3 v_0^\Gb b_{\Gb}^{\Ga} - v_0^3 v_0^{3,\Ga} \right)~, \\ 
    \rho_\mathrm{s}\dot{\bmv}_0 \cdot \bmn &= \rho_\mathrm{s}\left( v_{0,t}^3 + 2v_0^\Ga v_{0,\Ga}^3 + v_0^\Ga v_0^\Gb b_{\Gb \Ga} \right)~.
\end{align}

In direct surface theories, the stresses $N^{\Ga\Gb}$ and moments $M^{\Ga\Gb}$ are determined by two-dimensional constitutive models. In the $(2+\delta)$-dimensional theory, however, these quantities can be derived from three-dimensional constitutive models---as will be shown in Sec.~\ref{sec:const_models}. Furthermore, while we formally define the transverse shear stress $S^\Ga$ in terms of the three-dimensional stress, it cannot be determined from a constitutive model but is itself an unknown of the theory. This is a result of constraining the motion of the lipid membrane to satisfy Kirchhoff-Love kinematics, or, equivalently, to satisfy Eq.~\eqref{eq:velo_eulerian}. This constraint is enforced by so-called reactive stresses which cannot be determined a-priori. These reactive stresses enter the stress tensor components $\check{\sigma}^{i3} = \check{\sigma}^{3i}$ and consequently $S^\Ga$. Therefore, we use the angular momentum balance to derive an expression for $S^\Ga$, yielding (see part 2 \cite{omar2023BL} for details)
\begin{align}
    S^\Ga &= -M^{\gamma\alpha}_{;\gamma} - I^\Ga + P^\Ga~, \label{eq:2pd_Sa_short}
\end{align}
where
\begin{align}
    I^\Ga &=  \frac{\rho_{\mathrm{s}}\delta^2}{8} \left(  2H \dot{\bm{v}}_0 +   \ddot{\bm{n}}\right) \cdot \bma^\Ga~, \label{eq:Ia_def} \\
    P^\Ga &= \delta\left( \frac{1}{2}\left(\bm{t}^+ - \bm{t}^-\right) + \left\langle \left\llbracket \bm{\sigma}^T_\mathrm{M} \right\rrbracket \right\rangle^\mathcal{M} \bm{n} - \frac{\delta H}{4}\left(\bm{t}^+ + \bm{t}^- + \frac{1}{2}\left\llbracket \left\llbracket \bm{\sigma}^T_\mathrm{M} \right\rrbracket \right\rrbracket^\mathcal{M} \bm{n}\right)  - \delta\bm{f}_1 \right)\cdot \bma^\Ga~,  \label{eq:Pa_def}
\end{align}
with the acceleration of the normal vector given by
\begin{align}
    \ddot{\bmn} = \left( v^\Ga_{1,t} + v_0^\Gb v^\Ga_{1,\Gb} + v_1^\Gb v_{0:\Gb}^\Ga - v_1^\Gb v_0^3 b_\Gb^\Ga -  v_1^\Gb v^\Ga_{0,\Gb} \right) \bma_\Ga + \left( v_1^\Ga v_{0,\Ga}^3 + v_1^\Ga v_0^\Gb b_{\Gb \Ga} \right) \bmn ~. 
\end{align}
The terms
\begin{align}
    - \frac{\delta}{4} \left(  \rho_{\mathrm{s}}\delta H \dot{\bm{v}}_0 +  \frac{\rho_{\mathrm{s}}\delta}{2} \ddot{\bm{n}}  + 2\delta H\left(\frac{1}{2}\left(\bm{t}^+ + \bm{t}^-\right) + \frac{1}{2}\left\llbracket \left\llbracket \bm{\sigma}^T_\mathrm{M} \right\rrbracket \right\rrbracket^\mathcal{M} \bm{n}\right) \right)\cdot \bma^\Ga~, \label{eq:Sa_inplane_neglect}
\end{align}
appearing in Eq.~\eqref{eq:2pd_Sa_short}, can be shown to be negligible when substituted into the in-plane components of the linear momentum balance, Eq.~\eqref{eq:2pd_linmom_inplane}, as discussed in part 2 \cite{omar2023BL}. However, the terms in Eq.~\eqref{eq:Sa_inplane_neglect} cannot be neglected in the normal component of the linear momentum balance in Eq.~\eqref{eq:2pd_linmom_shape}. Furthermore, in the remainder of this article, we make the assumption that the first-order body term contribution vanishes, i.e. $\delta \bmf_1 \approx 0$. While this assumption is not essential, it simplifies some of the expressions derived in later sections.\textspace

Finally, we state the $(2+\delta)$-dimensional electromechanical boundary conditions corresponding to Eqs.~\eqref{eq:Spar_DEM} and~\eqref{eq:Spar_NEM}. Using the velocity expansion in Eq.~\eqref{eq:velo_chebT_expansion}, the $(2+\delta)$-dimensional velocity boundary conditions corresponding to Eq.~\eqref{eq:Spar_DEM} are given by
\begin{alignat}{2}
    v_0^\alpha &= \langle \bar{\bm{v}} \cdot \bm{a}^\alpha, \ChebT{0}{\Theta} \rangle~, \quad &&\forall \bm{x}_0 \in \partial \mathcal{S}_{||\mathrm{D}}^{\mathrm{EM}}~, \label{eq:BCSparD_inplane0}  \\
    v_0^3 &= \langle \bar{\bm{v}} \cdot \bm{n}, \ChebT{0}{\Theta} \rangle~, \quad &&\forall \bm{x}_0 \in \partial \mathcal{S}_{||\mathrm{D}}^{\mathrm{EM}}~, \label{eq:BCSparD_ooplane0}\\
    \delta v_1^\alpha &= 4\langle \bar{\bm{v}} \cdot \bm{a}^\alpha, \ChebT{1}{\Theta} \rangle~, \quad &&\forall\bm{x}_0 \in \partial \mathcal{S}_{||\mathrm{D}}^{\mathrm{EM}}~, \label{eq:BCSparD_inplane1}
\end{alignat}
where we defined $\partial\mathcal{S}_{0\mathrm{D}}^{\mathrm{EM}} = \mathcal{S}_{||\mathrm{D}}^{\mathrm{EM}} \cap \partial S_0$.
Similarly, with the binormal to the mid-surface boundary $\partial \mathcal{S}_0$, $\bm{\nu} = \check{\nu}_\Ga \bma^\Ga$, the $(2+\delta)$-dimensional in-plane traction boundary conditions corresponding to Eq.~\eqref{eq:Spar_NEM} are
\begin{alignat}{2} 
    \check{\nu}_\alpha \left( N^{\alpha\beta} + b_\gamma^\alpha M^{\gamma\beta}\right) &= \delta\langle \bma^\Gb \cdot \tilde{\bar{\bm{t}}}, \ChebT{0}{\Theta}\rangle~, \quad &&\forall \bm{x}_0 \in \partial \mathcal{S}_{0\mathrm{N}}^{\mathrm{EM}}~, \label{eq:t0_inplane_BC}\\
    \check{\nu}_\alpha \left( M^{\alpha\beta} + \frac{\delta^2}{16}b^\alpha_\gamma \left(2N^{\gamma\beta} + \delta\bmT_2^\Ga \cdot \bma^\Gb \right)\right) &= -\frac{\delta^2}{2}\langle \bma^\Gb \cdot \tilde{\bar{\bm{t}}}, \ChebT{1}{\Theta}\rangle~, \qquad &&\forall \bm{x}_0 \in \partial \mathcal{S}_{0\mathrm{N}}^{\mathrm{EM}}~, \label{eq:t1_inplane_BC}
\end{alignat}
where we defined $\partial \mathcal{S}_{0\mathrm{N}}^{\mathrm{EM}} = \mathcal{S}_{||\mathrm{N}}^{\mathrm{EM}} \cap \partial \mathcal{S}_0$ and 
\begin{align}
    \tilde{\bar{\bm{t}}} = \bar{\bm{t}} - \bm{\sigma}^T_{\mathrm{M}}\bm{\nu}~. \label{eq:t_tildebar}
\end{align}
Furthermore, we find the normal component of the traction boundary conditions in Eq.~\eqref{eq:Spar_NEM} to be
\begin{alignat}{2}
     \frac{\check{\nu}_\alpha}{2}\left( \bm{t}^+ - \bm{t}^-\right)\cdot\bm{a}^\alpha &= \langle \bmn\cdot\tilde{\bar{\bm{t}}}, \ChebT{0}{\Theta}\rangle~, \quad &&\forall \bm{x} \in \mathcal{S}_{||\mathrm{N}}^{\mathrm{EM}}~, \label{eq:t0_oop_BC}\\
    \frac{\check{\nu}_\alpha}{2} \left( \bm{t}^+ + \bm{t}^-\right)\cdot\bm{a}^\alpha &= 2 \langle \bmn \cdot \tilde{\bar{\bm{t}}}, \ChebT{1}{\Theta}\rangle~, \quad &&\forall \bm{x} \in \mathcal{S}_{||\mathrm{N}}^{\mathrm{EM}}~.\label{eq:t1_oop_BC}
\end{alignat}
Finally, we note that consistency with the three-dimensional starting point requires that only Eqs.~\eqref{eq:BCSparD_inplane0}--\eqref{eq:BCSparD_inplane1} or Eqs.~\eqref{eq:t0_oop_BC}--\eqref{eq:t1_oop_BC} are prescribed but no combinations thereof. This is in contrast to classical shell theories where, for example, moments and displacement can be specified together.

\paragraph{Order of Magnitude Assumptions}
To derive the $(2+\delta)$-dimensional balance laws and corresponding boundary conditions, we used a number of order of magnitude assumptions for the kinematics and stresses \cite{omar2023BL}. These assumptions, although rather weak, can and should be verified a-posteriori and are therefore briefly summarized here.\textspace

The first assumption is of a geometric nature and relates to the principal curvatures of the mid-surface, denoted by $\kappa_\Ga$ \cite{itskov2019tensor}. Specifically, we assume that the principal curvatures are small when compared to the thickness of the lipid membrane, such that 
\begin{align}
    \left( \delta \kappa_\alpha \right)^2 \ll 1~, \quad \alpha = 1,2~. \label{eq:ka_ll1}
\end{align}
Since $H=\frac{1}{2}\left(\kappa_1 + \kappa_2\right)$ and $K = \kappa_1\kappa_2$, Eq.~\eqref{eq:ka_ll1} also implies
\begin{align}
    \left( \delta H \right)^2 &\ll 1~, \label{eq:H2small}  \\
    \delta^2 \lvert K \rvert &\ll 1~. \label{eq:Ksmall}
\end{align}
While Eq.~\eqref{eq:ka_ll1} characterizes the out-of-plane or bending deformations of the mid-surface, we also introduce three in-plane characteristic length scales. Specifically, we assume there exist length scales $\ell_\mathrm{c}$, $\ell_\mathrm{s}$ and $\ell_\mathrm{v}$  describing changes in curvature, stresses and velocities along the in-plane directions, respectively. In addition, we assume the former two length scales satisfy
\begin{align}
    \frac{\ell_\mathrm{c}}{\ell_\mathrm{s}} &\not\ll 1~,\label{eq:lcls} \\
    \left( \frac{\delta}{\ell_\mathrm{c}} \right)^2 &\ll 1~, \label{eq:ellc_ll1} \\
    \left( \frac{\delta}{\ell_\mathrm{s}} \right)^2 &\ll 1~.\label{eq:ells_ll1}
\end{align}
implying $\ell_\mathrm{c}$ is not much smaller than $\ell_\mathrm{s}$, and that the two length scales $\ell_\mathrm{c}$ and $\ell_\mathrm{s}$ are large compared to the thickness of the membrane.
Note, however, that the assumptions in Eqs.~\eqref{eq:ellc_ll1} and~\eqref{eq:ells_ll1} could be relaxed without significantly changing the complexity of the theory (see also part 2 \cite{omar2023BL}). Lastly, the kinematics are simplified by assuming the zeroth- and first-order velocity components defined by Eq.~\eqref{eq:velo_eulerian} obey
\begin{align}
    \left(\frac{\delta v_1^\Ga}{v_0^\Gb}\right)^2 \ll 1~, \label{eq:va_1_magnitude}
\end{align}
which is, however, not guaranteed to be satisfied (cf. Eq.~\eqref{eq:v1_Eulerian}) and should therefore be verified a-posteriori. Next, we summarize the assumptions for the components of the stress tensor in Eq.~\eqref{eq:stress_tensor_expansion_bar}, which allow us to infer the order of magnitude of the stress vector expansion coefficients in Eqs.~\eqref{eq:Talpha_n_general} and~\eqref{eq:T3_n_general}, and, consequently, of the components $N^{\Ga\Gb}$, $M^{\Ga\Gb}$, and $S^\Ga$ in Eqs.~\eqref{eq:Nab_def},~\eqref{eq:Mab_def}, and~\eqref{eq:Sa_def}, respectively.\textspace

To introduce an assumption for the order of magnitude of the constitutively-determined components of the stress tensor, $\check{\sigma}^{\Ga\Gb}$, we assume they are given by a constitutive model that depends on a non-dimensional parameter $\lvert \mathcal{A} \rvert < 1$ such that
\begin{align}
    {\check{\sigma}}^{\alpha \beta}_k &= \mathcal{O}\leftR( \mathcal{A}^{\mathcal{Z}_k} \rightR)~, \label{eq:stress_assumption_1}
\end{align}
where the exponent $\mathcal{Z}_k$ depends on the order of the expansion coefficient $k$. We then assume that the exponents of the zeroth- and first-order terms are not larger than those of any higher-order terms, i.e. 
\begin{align}
    \mathcal{Z}_k \geq \mathcal{Z}_0 \quad \text{and} \quad \mathcal{Z}_k \geq \mathcal{Z}_1~, \qquad \forall k \geq 2~. \label{eq:stress_assumption_2}
\end{align}
If the constitutive model depends on multiple non-dimensional parameters, this assumption can be generalized by assuming that each parameter satisfies the above conditions.\textspace

Lastly, we require an order of magnitude assumption for the stress tensor components $\check{\sigma}^{i 3} = \check{\sigma}^{3 i}$ which are determined by the reactive stresses that enforce the linear velocity assumption in Eq.~\eqref{eq:velo_eulerian} (cf. part 2 \cite{omar2023BL}). For these, we assume the zeroth- and first-order coefficients are much larger than any higher-order coefficients, i.e.
\begin{align}
    \check{\sigma}_k^{i3} = \check{\sigma}_k^{3i} \gg \check{\sigma}_l^{i3} = \check{\sigma}_l^{3i}~, \quad k = 0,1~,~\forall l \geq 2~. \label{eq:OoM_penalty}
\end{align}
While this assumption is motivated considering purely elastic thin bodies in part 2 \cite{omar2023BL}, it cannot be justified a-priori for lipid membranes and its verification requires solving the constrained, three-dimensional problem. Nonetheless, this assumption allows us to significantly simplify the $(2+\delta)$-dimensional theory.

\section{Constitutive Models} \label{sec:const_models}
Capturing the viscous-elastic material behavior of lipid membranes requires a suitable choice of three-dimensional constitutive models. In this section, we propose elastic and viscous constitutive models relevant for lipid membranes and discuss how mid-surface area incompressibility can be enforced. Subsequently, we derive corresponding expressions for the in-plane stresses $N^{\Ga\Gb}$ and moments $M^{\Ga\Gb}$. To this end, we split $N^{\Ga\Gb}$ and $M^{\Ga\Gb}$ into elastic, viscous, and reactive contributions, 
\begin{align}
    N^{\Ga\Gb} &= N^{\Ga\Gb}_\mathrm{el} + \pi^{\Ga\Gb} + N^{\Ga\Gb}_\mathrm{r}~,\\
    M^{\Ga\Gb} &= M^{\Ga\Gb}_\mathrm{el} + M^{\Ga\Gb}_\mathrm{visc} + M^{\Ga\Gb}_\mathrm{r}~,
\end{align}
where $N^{\Ga\Gb}_\mathrm{el}$ and $M^{\Ga\Gb}_\mathrm{el}$, $\pi^{\Ga\Gb}$\footnote{This notation is used for consistency with previous studies \cite{rangamani2013interaction,sahu2017irreversible}.} and $M^{\Ga\Gb}_\mathrm{visc}$, and $N^{\Ga\Gb}_\mathrm{r}$ and $M^{\Ga\Gb}_\mathrm{r}$ are the stresses and moments arising from elastic and viscous constitutive models and mid-surface area incompressibility, respectively. In Secs.~\ref{sec:elastic}--\ref{sec:S0_incomp}, we derive suitable expressions for these stresses and moments that allow us to obtain the $(2+\delta)$-dimensional equations of motion for lipid membranes in Sec.~\ref{sec:EOMs}.

\subsection{Elastic Response}\label{sec:elastic}
The simplest choice of material model that describes the elastic response of lipid membranes---consistent with their in-plane fluidity---is one that penalizes volume changes. This can be modeled by the Helmholtz free energy per unit volume
\begin{align}
    w = \frac{k_\mathrm{c}}{J} \left(J-1\right)^2~, \label{eq:volume_Helmholtz}
\end{align}
where $k_\mathrm{c}$ is the compression modulus and $J$ is the pointwise volume change with respect to a stress-free reference configuration\footnote{The stress-free reference configuration does not necessarily coincide with the reference configuration considered in the derivation of the $(2+\delta)$-dimensional mass balance in part 2 \cite{omar2023BL}. The latter is marked by a constant density through the thickness and is not necessarily stress-free.\label{fn:refconfig}}.
This choice can be shown to yield the Cauchy stress \cite{coleman1963thermodynamics}
\begin{align}
    \bm{\sigma}_\mathrm{el}= 2 k_\mathrm{c} \left( J - 1 \right) \bm1~. \label{eq:const_elastic}
\end{align}
Since we showed in part 2 \cite{omar2023BL} that the Kirchhoff-Love assumption implies that only the in-plane stress components $\check{\sigma}^{\Ga\Gb}$ (cf. Eq.~\eqref{eq:sig_check_expansion}) can be prescribed by constitutive models, the three-dimensional metric tensor $\bm{1} = {\bmg}_i \otimes {\bmg}^i$ in Eq.~\eqref{eq:const_elastic} may be replaced by the metric tensor of the tangent space to the mid-surface $\bmi = {\bma}_\Ga \otimes {\bma}^\Ga$.\textspace

The volume change $J$ appearing in Eq.~\eqref{eq:const_elastic} can be expressed in terms of infinitesimal volume elements as $J = {\ddiff{v}}/{\ddiff{V}}$ with $\ddiff{v}$ and $\ddiff{V}$ denoting corresponding infinitesimal volume elements in the current and reference configurations, respectively. These volume elements can be further expressed in terms of their heights $\ddiff{\mathrm{w}}$ and $\ddiff{W}$, and areas parallel to the mid-surface, $\ddiff{a}$ and $\ddiff{A}$, as $\ddiff{v} = \ddiff{\mathrm{w}} \ddiff{a}$ and $\ddiff{V} = \ddiff{W} \ddiff{A}$. However, the Kirchhoff-Love assumption implies $\ddiff{\mathrm{w}} = \ddiff{W}$ such that $J$ can be written in terms of area changes as\footnote{Technically, the expressions for $J$ should be first written in terms of the Lagrangian parametrization. However, the Kirchhoff-Love assumptions and the choice of the form of the position vector in Eq.~\eqref{eq:xform} imply that we obtain the same result when starting with the Eulerian parametrization.} 
\begin{align}
    J = \frac{\ddiff{a}}{\ddiff{A}}~, \label{eq:J3D}
\end{align}
The area elements $\ddiff{a}$ and $\ddiff{A}$ can be related to the infinitesimal area elements on the mid-surfaces of the current and reference configurations, denoted by $\ddiff{a}_0$ and $\ddiff{A}_0$, respectively. By using expressions for the area change through the thickness under the assumption of Kirchhoff-Love kinematics \cite{chien1944intrinsic,green1950equilibrium,naghdi1962foundations,song2016consistent}, Eq.~\eqref{eq:J3D} takes the form 
\begin{align}
    J = J_0 \left( \frac{ 1 - \delta H \Theta + \frac{\delta^2}{4}K \Theta^2}{ 1 - \delta C \Theta + \frac{\delta^2}{4}G \Theta^2}\right)~, \label{eq:Jrational}
\end{align}
where $J_0$ is the relative area change of the mid-surface, i.e.
\begin{align}
    J_0 = \frac{\ddiff{a}_0}{\ddiff{A}_0}~,
\end{align}
and $C$ and $G$ are the mean and Gaussian curvatures in the reference configuration, respectively. As in the current configuration, we assume that the principal curvatures $\kappa_{0\Ga}$ in the reference configuration are small compared to the thickness, i.e.
\begin{align}
    \left(\delta \kappa_{0\alpha}\right)^2 \ll 1~, \label{eq:dk02_small}
\end{align}
which further implies
\begin{align}
    \left(\delta C\right)^2 &\ll 1~, \label{eq:C2small} \\
    \delta^2 \lvert G \rvert & \ll 1~. \label{eq:Gsmall}
\end{align}

Using Eqs.~\eqref{eq:const_elastic} and~\eqref{eq:Jrational} and the assumptions in Eqs.~\eqref{eq:H2small},~\eqref{eq:Ksmall},~\eqref{eq:C2small}, and~\eqref{eq:Gsmall}, the elastic contributions to the stresses and moments in Eqs.~\eqref{eq:Nab_def} and~\eqref{eq:Mab_def} become (see SM, Sec.~3.1)
\begin{align}
    N^{\Ga\Gb}_{\mathrm{el}} &\approx 2\bar{k}_\mathrm{c} \left(J_0 -1\right) a^{\Ga\Gb} + J_0 k_\mathrm{b} \left( -\left(H-C\right)\left( 2C a^{\Ga\Gb} + b^{\Ga\Gb}\right) + \frac{1}{2}\left(K-G\right)a^{\Ga\Gb} \right)~, \label{eq:Nab_el} \\
    M^{\Ga\Gb}_\mathrm{el} &\approx J_0 k_\mathrm{b} \left(H-C\right)a^{\Ga\Gb}-\frac{k_\mathrm{b}}{2}\left(J_0 -1\right)b^{\Ga\Gb} ~,\label{eq:Mab_el}
\end{align}
where we defined the effective compression modulus $\bar{k}_\mathrm{c}$ and bending rigidity $k_\mathrm{b}$ as
\begin{align}
    \bar{k}_\mathrm{c} &= \delta {k}_\mathrm{c}~, \label{eq:barkc_def}\\
    k_\mathrm{b} &= \frac{{k}_\mathrm{c}\delta^3}{2}~. \label{eq:kb_def}
\end{align}
Combining Eqs.~\eqref{eq:barkc_def} and~\eqref{eq:kb_def} yields a relation between the bending and compression modulus,
\begin{align}
    k_\mathrm{b} = \frac{\bar{k}_\mathrm{c} \delta^2}{2}~, \label{eq:kb_kc_rel}
\end{align}
consistent with the result obtained by Evans \cite{evans1974bending}. \textspace

To compare Eqs.~\eqref{eq:Nab_el} and~\eqref{eq:Mab_el} to strict surface theories \cite{rangamani2013interaction,sahu2017irreversible}, recall that the stresses and moments resulting from the Canham-Helfrich-Evans model \cite{canham:1970,helfrich1973elastic,evans:1974} are given by
\begin{align}
    N^{\Ga\Gb}_{2\mathrm{d},\mathrm{el}} &= 2\bar{k}_\mathrm{c} \left(J_0 -1\right) a^{\Ga\Gb} + k_\mathrm{b} \left( \left(H-C\right)^2 a^{\Ga\Gb} - \left(H-C\right) b^{\Ga\Gb} \right)~, \label{eq:Nab_2del} \\
    M^{\Ga\Gb}_{2\mathrm{d},\mathrm{el}} &= k_\mathrm{b} \left(H-C\right)a^{\Ga\Gb} + k_\mathrm{g}\left(2Ha^{\Ga\Gb} - b^{\Ga\Gb}\right)~.\label{eq:Mab_2del}
\end{align}
Comparison of the $(2+\delta)$-dimensional stress in Eq.~\eqref{eq:Nab_el} to the two-dimensional stress in Eq.~\eqref{eq:Nab_2del} shows that the mid-surface area stretch contributions are identical in both theories. However, while both theories also yield nonlinear bending contributions, their forms differ between the two theories. Furthermore, both the Gaussian and spontaneous Gaussian curvatures appear in Eq.~\eqref{eq:Nab_el} but not in Eq.~\eqref{eq:Nab_2del}. When comparing the elastic moments in Eqs.~\eqref{eq:Mab_el} and~\eqref{eq:Mab_2del}, we find that the first terms agree up to a scaling by $J_0$. However, the moments of the Canham-Helfrich-Evans model contain a Gaussian bending rigidity contribution that does not have an equivalent term in Eq.~\eqref{eq:Mab_el}. In contrast, the $(2+\delta)$-dimensional theory includes a coupling term between the mid-surface area stretch and curvature tensor that is not captured by strict surface theories. Lastly, we note that the scaling of the bending rigidity by the mid-surface stretch $J_0$ in Eqs.~\eqref{eq:Nab_el} and~\eqref{eq:Mab_el} is negligible since $J_0 \approx 1$ for lipid membranes \cite{evans1976elastic,nichol1996tensile}.\textspace

\subsection{Viscous Response}\label{sec:viscous}
To capture the in-plane fluid nature of lipid membranes, we assume the three-dimensional viscous material response is Newtonian such that the associated Cauchy stress is given by \cite{gurtin2010mechanics}
\begin{align}
    \bm{\sigma}_\mathrm{visc} = 2\mu \, \bm{D} + \omega \divv{\bm{v}} \bm1~. \label{eq:const_visc}
\end{align}
Here, $\mu$ and $\omega$ are the three-dimensional shear and bulk viscosities, respectively, and $\bmD$ is the symmetric part of the velocity gradient,
\begin{align}
    \bmD = \frac{1}{2} \left( \grad{\bmv} + \grad{\bmv}^T \right)~. 
\end{align}
As with the elastic stress in Eq.~\eqref{eq:const_elastic}, we could replace $\bmD$ and $\bm1$ by $\bmi\bmD\bmi$ and $\bmi$, respectively, given that we can only prescribe the in-plane components of the stress tensor due to the Kirchhoff-Love assumptions \cite{omar2023BL}. Equation~\eqref{eq:Nab_def} then yields the viscous contribution to $N^{\Ga\Gb}$ corresponding to Eq.~\eqref{eq:const_visc} as (see SM, Sec.~2.2)
\begin{align}
    \pi^{\Ga\Gb} &= \zeta \left( w^\Ga_{\,.\,\Gg}a^{\Gg\Gb} + w^\Gb_{\,.\,\Gg} a^{\Gg\Ga} \right) + \bar{\omega} \left( v^\Ga_{0:\Ga} - 2v_0^3 H \right)~, \label{eq:Nab_visc}
\end{align}
with
\begin{align}
    w^\Ga_{\,.\,\Gb} &= v^\Ga_{0:\Gb} - v_0^3 b^\Ga_\Gb~,
\end{align}
and the effective shear and bulk viscosities respectively defined as
\begin{align}
    \zeta &= \delta \mu~,\\
    \bar{\omega} &= \delta \omega~.
\end{align}
Note that in obtaining Eq.~\eqref{eq:Nab_visc}, we used the assumptions in Eqs.~\eqref{eq:H2small}--\eqref{eq:va_1_magnitude}. Furthermore, Eq.~\eqref{eq:Nab_visc} is in exact agreement with the viscous contributions of strict surface theories \cite{aris2012vectors,rangamani2013interaction,sahu2017irreversible}, thus warranting no further discussion.\textspace

The viscous stress in Eq.~\eqref{eq:const_visc} also yields a viscous contribution to the moments in Eq.~\eqref{eq:Mab_def}, namely (see SM, Sec.~2.2)
\begin{align}
    M^{\Ga\Gb}_\mathrm{visc} =& \: -\frac{\delta^3\mu}{4} \left( w^\Gb_{\,.\,\Gg} b^{\Ga\Gg} + \frac{1}{2}w^\mu_{\,.\,\Gg}\left( a^{\Gg\Gb} b^{\Ga}_\mu + b^{\Gg\Gb} \delta^\Ga_\mu \right)  + v^\Gb_{1:\Gg} \, a^{\Ga\Gg} + v^\Ga_{1:\Gg} \, a^{\Gg\Gb} \right) \: \nonumber \\
    & \: -\frac{\omega\delta^3}{16}\left( v_{1:\Gg}^\Gg a^{\Ga\Gb} + b^\Gg_\lambda w^{\lambda}_{\,.\,\Gg} a^{\Ga\Gb} + w^\Gg_{\,.\,\Gg} b^{\Ga\Gb}\right)~. 
\end{align}
However, it is customary to neglect viscous moments for lipid membranes as they correspond to effects at length scales comparable to the thickness and therefore, we set
\begin{align}
    M^{\Ga\Gb}_\mathrm{visc} = 0~. \label{eq:Mab_visc}
\end{align}

\subsection{Mid-surface area incompressibility} \label{sec:S0_incomp}
Lipid membranes typically only sustain a mid-surface stretch of up to $2\text{--}4\%$ \cite{evans1976elastic,nichol1996tensile}, motivating their treatment as mid-surface area incompressible materials. This assumption significantly simplifies solving the equations of motion analytically. However, constraining the mid-surface to be incompressible requires the introduction of additional reactive stresses \cite{carlson2004geometrically,rajagopal2005nature}. To this end, the area incompressibility constraint can be stated in terms of the mid-surface stretch,
\begin{align}
    J_0 = 1~, \label{eq:J0_1}
\end{align}
or in terms of the rate of change of the mid-surface density (cf. Eq.~\eqref{eq:2pd_mass_balance}),
\begin{align}
    v_{0:\Ga}^\Ga- 2  v H = 0~. \label{eq:divsv_0}
\end{align}

The reactive stresses associated with the constraints in Eqs.~\eqref{eq:J0_1} and~\eqref{eq:divsv_0} can be written in terms of the unknown coefficients $\tilde{\lambda}_k$ as (see SM, Sec.~2.3)
\begin{align}
    \bm{\sigma}_\mathrm{r} = \tilde{\lambda} \bmi = \left( \sum_{k=0}^\infty \tilde{\lambda}_k \, \Theta^k \right) \bma_\Ga \otimes \bma^\Ga~. \label{eq:sigr_i}
\end{align}
This gives rise to the reactive in-plane stresses and moments (see SM, Sec.~2.3),  
\begin{align}
    N^{\Ga\Gb}_\mathrm{r} &= \left(\frac{\delta}{2} 
     \sum_{m=0}^\infty \tilde{\lambda}_{2m} \Ga_{(2m)0} \right) a^{\Ga\Gb}~, \label{eq:Nab_r_red_1} \\
    M^{\Ga\Gb}_\mathrm{r} &= \left(-\frac{\delta^3}{8} 
    \sum_{m=0}^\infty \tilde{\lambda}_{2m} \Ga_{(2m+1) 1} \right) b^{\Ga\Gb}~. \label{eq:Mab_r_red_1}
\end{align}
In deriving Eqs.~\eqref{eq:Nab_r_red_1} and \eqref{eq:Mab_r_red_1}, we assumed that all coefficients $\tilde{\lambda}_k$ that are odd about the mid-surface vanish identically. 
This is motivated by the understanding that the reactive stresses balance the constitutively-determined stresses precisely such that the mid-surface stretch vanishes. However, to affect the mid-surface stretch, the reactive stresses must be non-vanishing at the mid-surface. Thus, we may assume that only the even terms of the sum in Eq.~\eqref{eq:sigr_i} are non-zero.
Furthermore, we later show that the moments $M^{\Ga\Gb}_\mathrm{r}$ in Eq.~\eqref{eq:Mab_r_red_1} are negligible in the equations of motion and boundary conditions (see also SM, Secs.~3 and 4). This suggests the definition of the effective surface tension $\lambda$ as
\begin{align}
    \lambda = \frac{\delta}{2} 
     \sum_{m=0}^\infty \tilde{\lambda}_{2m} \Ga_{(2m)0}~, \label{eq:lambda_def} 
\end{align}
such that the in-plane stresses in Eq.~\eqref{eq:Nab_r_red_1} can be written in the form
\begin{align}
    N^{\Ga\Gb}_\mathrm{r} = \lambda a^{\Ga\Gb}~, \label{eq:Nab_r_lambda}
\end{align}
consistent with direct surface theories of lipid membranes \cite{rangamani2013interaction, sahu2017irreversible}. The surface tension $\lambda$ can now be treated as an effective unknown that is found by solving the equations of motion along with the incompressibility constraint in Eq.~\eqref{eq:divsv_0}.

\section{$(2+\delta)$-Dimensional Equations of Motion}\label{sec:EOMs}

We now use the elastic, viscous, and reactive stresses and moments derived in Secs.~\ref{sec:elastic}--\ref{sec:S0_incomp} to obtain the $(2+\delta)$-dimensional equations of motion describing the electromechanics of lipid membranes. As before, only the main results are presented while the algebraic details are shown in Secs.~3 and 4 of the SM.

\subsection{Compressible Equations of Motion} \label{sec:comp_EOM}
We begin by deriving the equations of motion for the mid-plane compressible case. To obtain the in-plane equations, we substitute the elastic and viscous stresses and moments in Eqs.~\eqref{eq:Nab_el},~\eqref{eq:Mab_el},~\eqref{eq:Nab_visc}, and~\eqref{eq:Mab_visc} and the expression for $S^\Ga$ in Eq.~\eqref{eq:2pd_Sa_short} into 
Eq.~\eqref{eq:2pd_linmom_inplane}. By using the assumption of small curvatures in the current and reference confugurations in Eqs.~\eqref{eq:H2small}, \eqref{eq:Ksmall}, \eqref{eq:C2small} and~\eqref{eq:Gsmall}, the assumption $\ell_\mathrm{c}/\ell_\mathrm{s} \not \ll 1$ in  Eq.~\eqref{eq:lcls}, and the relation between the membrane bulk modulus and bending rigidity in Eq.~\eqref{eq:kb_kc_rel} (see Sec.~\ref{sec:elastic} and SM, Sec.~3), we obtain
\begin{align}
     \rho_\mathrm{s}\dot{\bmv}_0 \cdot \bma^\Ga &= 2\bar{k}_\mathrm{c} J_{0,\Gb} a^{\Ga\Gb} + J_0 k_\mathrm{b} \left( -2\left(C\left(H-C\right)\right)_{,\Gb}  + \frac{1}{2}\left(K-G\right)_{,\Gb} \right)a^{\Ga\Gb} + \pi^{\Gb\Ga}_{;\Gb}  \nonumber \\ 
    &\hspace{3cm} + \bm{f}_\mathrm{s} \cdot \bm{a}^\Ga + \left( \left(\bm{t}^+ + \bm{t}^-\right) + \left\llbracket \left\llbracket \bm{\sigma}^T_\mathrm{M} \right\rrbracket \right\rrbracket^\mathcal{M} \bm{n} \right) \cdot \bm{a}^\Ga 
    \nonumber \\
    &\hspace{4cm}- \delta\left( 2H\delta_\Gg^\Ga + b_\Gg^\Ga\right)\left(\frac{1}{2}\left(\bm{t}^+ - \bm{t}^-\right) + \left\langle \left\llbracket \bm{\sigma}^T_\mathrm{M} \right\rrbracket \right\rangle^\mathcal{M} \bm{n}\right) \cdot \bm{a}^\Gg~, \label{eq:2pd_inplane_comp}
\end{align}
where $\delta^\Ga_\Gg$ is the Kronecker delta and $\pi^{\Gb\Ga}_{;\Gb}$ denotes the covariant derivative of $\pi^{\Gb\Ga}$.\textspace

Similarly, substituting the elastic and viscous responses into the normal component of the linear momentum balance in Eq.~\eqref{eq:2pd_linmom_shape}, and using the assumption of small curvatures and the assumption that changes in curvature and stresses occur over length scales much larger than the thickness (see Eqs.~\eqref{eq:ellc_ll1} and~\eqref{eq:ells_ll1}), we find
\begin{align}
    \rho_\mathrm{s}\dot{\bmv}_0\cdot \bmn + I^\Ga_{:\Ga} &=  \: 4\bar{k}_\mathrm{c}\left(J_0 - 1\right) H + J_0 k_\mathrm{b}\left( - 2\left(H-C\right)\left(2H^2 + 2C H - K\right) + \left(K-G\right)H \right) \nonumber \\
    & \hspace{0.07\textwidth} - J_0 {k}_\mathrm{b}\Delta_\mathrm{s} \leftR(H-C\rightR) + \pi^{\alpha\beta}b_{\alpha\beta} \:
    +  P^\Ga_{:\Ga} + \left( \bm{t}^+ + \bm{t}^- + \left\llbracket \left\llbracket \bm{\sigma}^T_\mathrm{M} \right\rrbracket \right\rrbracket^\mathcal{M} \bm{n}\right) \cdot \bm{n} \nonumber \\ 
    & \hspace{0.25\textwidth}- 2\delta H \left(\frac{1}{2}\left(\bm{t}^+ - \bm{t}^-\right) + \left\langle \left\llbracket \bm{\sigma}^T_\mathrm{M} \right\rrbracket \right\rangle^\mathcal{M} \bm{n}\right)\cdot\bm{n} + \bm{f}_{\mathrm{s}} \cdot \bmn~, \label{eq:2pd_shape_comp}
\end{align}
where $\Delta_s\leftR(H-C\rightR) = \left( \left(H-C\right)_{,\Ga}\right)_{:\Gb} a^{\Ga\Gb}$ is the surface Laplacian. Equations~\eqref{eq:2pd_inplane_comp} and~\eqref{eq:2pd_shape_comp} together with the membrane mass balance in Eq.~\eqref{eq:2pd_mass_balance} constitute the compressible equations of motion of the $(2+\delta)$-dimensional theory of the electromechanics of lipid membranes.\textspace

We can compare the constitutive contributions to the $(2+\delta)$-dimensional equations of motion to those from strict surface theories that use the elastic Canham-Helfrich-Evans and in-plane Newtonian fluid models \cite{arroyo2009relaxation,rangamani2013interaction,sahu2017irreversible}. First, recall from Sec.~\ref{sec:viscous} that the viscous stresses and moments coincide with their counterparts of strict surface theories and consequently, the viscous contributions also agree in the in-plane and shape equations in Eqs.~\eqref{eq:2pd_inplane_comp} and \eqref{eq:2pd_shape_comp}. Furthermore, all area dilation terms involving $\bar{k}_\mathrm{c}$, and the surface Laplacian term in the shape equation are also found in strict surface theories. In contrast, the nonlinear bending terms involving $k_\mathrm{b}$ in the in-plane and shape equations differ from strict surface theories, as summarized in Table~\ref{tab:2pd_strict_comp}. Lastly, the bending rigidity $k_\mathrm{b}$ carries a prefactor $J_0$ in the $(2+\delta)$-dimensional theory, representing a curvature-stretch coupling not present in strict surface theories. However, as discussed in Sec.~\ref{sec:elastic}, this prefactor is likely negligible since $J_0 \approx 1$ for lipid membranes. 

\begin{table}[t]
    \centering
\begin{tabular}{c|c|c}
\toprule
&$(2+\delta)$\text{-dimensional theory} &  \text{Canham-Helfrich-Evans} \\
\toprule
Eq.~\eqref{eq:2pd_inplane_comp} & $-2\left(C\left(H-C\right)\right)_{,\Gb} + \frac{1}{2}\left(K-G\right)_{,\Gb} $ & $-2\left(H-C\right)C_{,\Gb}$  \\[3pt]\midrule
Eq.~\eqref{eq:2pd_shape_comp} & $- 2\left(H-C\right)\left(2H^2 + 2HC - K\right) + \left(K-G\right)H$ & $-2\left(H-C\right)\left(H^2 + H C - K\right)$ \\
\bottomrule
\end{tabular}
\caption{Comparison of the bending contributions in the in-plane and shape equations in Eqs.~\eqref{eq:2pd_inplane_comp} and~\eqref{eq:2pd_shape_comp} and strict surface theories using the Canham-Helfrich-Evans model \cite{arroyo2009relaxation,rangamani2013interaction,sahu2017irreversible}. } 
\label{tab:2pd_strict_comp}
\end{table}

\subsection{Incompressible Equations of Motion} \label{sec:incomp_EOM}
To derive the equations of motion for the incompressible case, we additionally account for the constraint stresses and moments in Eqs.~\eqref{eq:Nab_r_lambda} and~\eqref{eq:Mab_r_red_1}, respectively, and set $J_0 = 1$ and $\dot{\rho}_\mathrm{s} = 0$. This reduces the mass balance in Eq.~\eqref{eq:2pd_mass_balance} to the mid-surface area incompressibility condition,
\begin{align}
    v_{0:\Ga}^\Ga- 2  v_0^3 H = 0~.  \label{eq:2pd_mass_balance_comp}
\end{align}
The in-plane and shape equations, respectively, simplify to
\begin{align}
     \rho_\mathrm{s}\dot{\bmv}_0 \cdot \bma^\Ga &= \lambda_{,\Gb} a^{\Ga\Gb} + k_\mathrm{b} \left( -2\left(C\left(H-C\right)\right)_{,\Gb} + \frac{1}{2}\left(K-G\right)_{,\Gb}  \right)a^{\Ga\Gb} + \pi^{\Gb\Ga}_{;\Gb}  \nonumber \\ 
    & \hspace{2.5cm} + \bm{f}_\mathrm{s} \cdot \bm{a}^\Ga + \left( \left(\bm{t}^+ + \bm{t}^-\right) + \left\llbracket \left\llbracket \bm{\sigma}^T_\mathrm{M} \right\rrbracket \right\rrbracket^\mathcal{M} \bm{n} \right) \cdot \bm{a}^\Ga 
    \nonumber \\ 
    & \hspace{3.5cm} - \delta\left( 2H\delta_\Gg^\Ga + b_\Gg^\Ga\right)\left(\frac{1}{2}\left(\bm{t}^+ - \bm{t}^-\right) + \left\langle \left\llbracket \bm{\sigma}^T_\mathrm{M} \right\rrbracket \right\rangle^\mathcal{M} \bm{n}\right) \cdot \bm{a}^\Gg~, \label{eq:2pd_inplane_incomp}
\end{align}
and
\begin{align}
    \rho_\mathrm{s}\dot{\bmv}_0\cdot \bmn + I^\Ga_{:\Ga} &= 2
    \lambda H + k_\mathrm{b}\left( - 2\left(H-C\right)\left(2H^2 + 2C H - K\right) + \left(K-G\right)H \right) \nonumber \\
    & \hspace{0.07\textwidth} - {k}_\mathrm{b}\Delta_\mathrm{s} \left(H-C\right)  + \pi^{\alpha\beta}b_{\alpha\beta} +   P^\Ga_{:\Ga} + \left( \bm{t}^+ + \bm{t}^- + \left\llbracket \left\llbracket \bm{\sigma}^T_\mathrm{M} \right\rrbracket \right\rrbracket^\mathcal{M} \bm{n}\right) \cdot \bm{n} \nonumber \\
    &  \hspace{0.223\linewidth} - 2\delta H \left(\frac{1}{2}\left(\bm{t}^+ - \bm{t}^-\right) + \left\langle \left\llbracket \bm{\sigma}^T_\mathrm{M} \right\rrbracket \right\rangle^\mathcal{M} \bm{n}\right)\cdot\bm{n} + \bm{f}_{\mathrm{s}} \cdot \bmn~, \label{eq:2pd_shape_incomp}
\end{align}
where we have used the same assumptions as in Sec.~\ref{sec:comp_EOM}. Comparison of these equations with the compressible equations of motion, Eqs.~\eqref{eq:2pd_inplane_comp} and~\eqref{eq:2pd_shape_comp}, 
reveals little difference. However, the area dilation contributions are now replaced by tension-related terms that match their counterparts in strict surface theories \cite{arroyo2009relaxation,rangamani2013interaction,sahu2017irreversible}.

\subsection{Boundary Conditions}
Using the constitutive and reactive stresses and moments from Sec.~\ref{sec:const_models}, we now derive the corresponding boundary conditions. To this end, first note that the out-of-plane traction boundary conditions in Eqs.~\eqref{eq:t0_oop_BC} and~\eqref{eq:t1_oop_BC} and the Dirichlet boundary conditions in Eqs.~\eqref{eq:BCSparD_inplane0}--\eqref{eq:BCSparD_inplane1} are independent of the choice of constitutive models. Thus, we only need to consider the in-plane traction boundary conditions in Eqs.~\eqref{eq:t0_inplane_BC} and~\eqref{eq:t1_inplane_BC} for the reminder of this section.

\paragraph{Compressible Boundary Conditions} 
Beginning with the compressible case, using the viscous and elastic stresses, moments and expressions for $\bmT_2^\Ga$ derived in Secs.~\ref{sec:elastic}, and applying the assumptions of small curvatures in Eqs.~\eqref{eq:ka_ll1}--\eqref{eq:Ksmall},~\eqref{eq:C2small} and~\eqref{eq:Gsmall} as well as Eq.~\eqref{eq:kb_kc_rel}, the in-plane traction boundary conditions in Eqs.~\eqref{eq:t0_inplane_BC} and~\eqref{eq:t1_inplane_BC} yield
\begin{align}
    \check{\nu}^\Gb \left( 2\bar{k}_\mathrm{c}\left(J_0-1\right) + J_0 k_\mathrm{b}\left( -2C \left(H-C\right) + \frac{1}{2}\left(K-G\right)\right) \right) + \check{\nu}_\Ga \pi^{\Ga\Gb} &= \delta\langle \bma^\Gb \cdot \tilde{\bar{\bm{t}}}, \ChebT{0}{\Theta}\rangle~, \label{eq:t0_inplane_BC_comp} \\
    \check{\nu}_\alpha \left( J_0 k_\mathrm{b}\left(H-C\right)a^{\Ga\Gb} + \frac{\delta^2}{8}b^{\Ga}_\Gg \pi^{\Gg\Gb} \right) &= -\frac{\delta^2}{4}\langle \bma^\Gb \cdot \tilde{\bar{\bm{t}}}, \ChebT{1}{\Theta}\rangle~. \label{eq:t1_inplane_BC_comp}
\end{align}

\begin{table}[t]
    \centering
\begin{tabular}{c|c}
\toprule
$(2+\delta)$\text{-dimensional theory} &  \text{Canham-Helfrich-Evans} \\
\toprule
$J_0 k_\mathrm{b}\check{\nu}^\Gb\left(-2C \left(H-C\right) + \frac{1}{2}\left(K-G\right)\right)$ & $k_\mathrm{b}\left(\left(H-C\right)^2\check{\nu}^\Gb - \left(H-C\right) b^{\Ga\Gb} \check{\nu}_\Ga\right) - k_\mathrm{g}\check{\nu}_\alpha b^{\Ga\Gg}b^{\mu\Gb}\tau_\Gg \tau_\mu  $ \\
\bottomrule
\end{tabular}
\caption{Comparison of the bending terms in the in-plane traction boundary condition in Eq.~\eqref{eq:t0_inplane_BC_comp} to those derived for strict surface theories \cite{arroyo2009relaxation,rangamani2013interaction,sahu2017irreversible} (see SM, Sec.~5). In the Canham-Helfrich-Evans theory, $k_\mathrm{g}$ denotes the Gaussian bending rigidity and $\tau_\Gg$ are the components of the tangent vector on $\partial \mathcal{S}_0$, $\bm{\tau} = \bmn \times \bm{\nu}$.} 
\label{tab:2pd_strict_comp_bct0}
\end{table}

Comparison of the zeroth-order, in-plane boundary conditions in Eq.~\eqref{eq:t0_inplane_BC_comp} with those of strict surface theories \cite{rangamani2013interaction,sahu2017irreversible} again shows agreement for the area dilation and viscous terms. The terms associated with the bending rigidity $k_\mathrm{b}$, however, differ from those of strict surface theories as can be seen in Table~\ref{tab:2pd_strict_comp_bct0}. In addition, the Canham-Helfrich-Evans model contains a Gaussian bending rigidity term not present in the $(2+\delta)$-dimensional theory. 
When comparing the first-order boundary condition in Eq.~\eqref{eq:t1_inplane_BC_comp} with the moment boundary conditions of strict surface theories, we find the same bending related term $k_\mathrm{b}\left(H-C\right)a^{\Ga\Gb}$. As in the zeroth-order case, however, the moment boundary conditions of strict surface theories also contain a term related to the Gaussian bending rigidity not found in the $(2+\delta)$-dimensional theory. In contrast, Eq.~\eqref{eq:t1_inplane_BC_comp} contains a viscous stress term not present in strict surface theories. Lastly, as with the equations of motion, the bending rigidity carries the mid-surface stretch as a prefactor in the $(2+\delta)$-dimensional theory in Eqs.~\eqref{eq:t0_inplane_BC_comp} and \eqref{eq:t1_inplane_BC_comp}. This prefactor is not found in strict surface theories but is negligible for lipid membranes.

\paragraph{Incompressible boundary conditions}
The boundary conditions for the incompressible case are obtained by additionally accounting for the reactive stresses in the boundary conditions in Eqs.~\eqref{eq:t0_inplane_BC} and~\eqref{eq:t1_inplane_BC} and setting $J_0 = 1$, yielding
\begin{align} 
    \check{\nu}^\Gb \left( \lambda + k_\mathrm{b}\left( -2C \left(H-C\right) + \frac{1}{2}\left(K-G\right)\right) \right) + \check{\nu}_\Ga \pi^{\Ga\Gb} &= \delta\langle \bma^\Gb \cdot \tilde{\bar{\bm{t}}}, \ChebT{0}{\Theta}\rangle~,\label{eq:t0_inplane_BC_incomp}\\
    \check{\nu}_\alpha \left(k_\mathrm{b}\left(H-C\right)a^{\Ga\Gb} + \frac{\delta^2}{8}b^{\Ga}_\Gg \pi^{\Gg\Gb} \right) &= -\frac{\delta^2}{2}\langle \bma^\Gb \cdot \tilde{\bar{\bm{t}}}, \ChebT{1}{\Theta}\rangle~. \label{eq:t1_inplane_BC_incomp} 
\end{align}
Note that the reactive stress contribution vanishes identically in the first-order boundary condition. This result and the tension term in Eq.~\eqref{eq:t0_inplane_BC_incomp} are consistent with strict surface theories \cite{rangamani2013interaction,sahu2017irreversible}.

\section{Equations of Motion for a Lipid Membrane in Contact with an Electrolyte} \label{sec:EOMElectrolyte}
To summarize the results obtained in this sequence of articles including earlier parts 1 and 2 \cite{omar2024ES,omar2023BL}, we describe the set of equations modeling a lipid membrane in contact with an electrolyte. To simplify our results, we assume that there is no spontaneous curvature ($C = G = 0$), the membrane is incompressible and that all inertial terms are negligible. In the following, we state all coupling conditions and boundary conditions of the membrane but omit the well-known boundary conditions of the bulk fluid for brevity. 

\paragraph{Bulk domains}
Suppose each of the bulk domains on either side of the membrane is an electrolyte containing $N$ species with concentrations $c_i^\pm$ and valency $z_i$.
Here, we assume that the electrolytes are linear dielectric materials. By further positing that the electric field evolves sufficiently slow such that the electrostatic assumptions hold (see SM of part 1 \cite{omar2024ES}), the electric potential in the bulk, $ {\phi}_{\mathcal{B}^\pm}$, is governed by Gauss' law \cite{melcher1981continuum,omar2024ES}, i.e.
\begin{align}
    \Delta {\phi}_{\mathcal{B}^\pm} &= -\frac{1}{\varepsilon_{\mathcal{B}}} \sum_{i=1}^N c_i^\pm z_i e~.
\end{align}
Here, $e$ is the elementary charge and $\varepsilon_{\mathcal{B}} = \varepsilon_{\mathcal{B}}^+ = \varepsilon_{\mathcal{B}}^-$ is the bulk permittivity, which we assume equal on both sides of the membrane.\textspace

Next, the species mass balances of the electrolytes are given by
\begin{align}
    \frac{\ddiff c_i^\pm}{\ddiff t} + c_i^\pm \divv{\bmu^\pm} = - \divv{\bmJ_i^\pm}~, \quad i = 1,..., N~, \label{eq:speciesMass}
\end{align}
where $\bmu^\pm$ is the bulk velocity and $\bmJ_i^\pm$ is the diffusive flux of species $i$. Assuming infinite dilution and an ideal solution, the constitutive laws for the diffusive fluxes are given as~\cite{fong2020transport}
\begin{align}
    \bmJ_i^\pm = -D_i \grad{c_i^\pm} - \frac{D_i z_i F}{RT} c_i^\pm \grad{\phi_{\mathcal{B}^\pm}}~. \label{eq:PNPFlux}
\end{align}
where $D_i$ is the diffusion coefficient of species $i$, $F$ is Faraday's constant, $R$ is the ideal gas constant, and $T$ is temperature. The first term on the right-hand side of Eq.~\eqref{eq:PNPFlux} is Fick's diffusion law while the second term describes electromigration. Combining Eqs.~\eqref{eq:speciesMass} and \eqref{eq:PNPFlux} leads to the well-known Poisson-Nernst-Planck equation.\textspace

Finally, we require the linear momentum balance to determine the velocities $\bmu^\pm$, which is given by 
\begin{align}
    \bm0 = \divv{\bbar{\bm{\sigma}}_{\mathcal{B}^\pm}}~,
\end{align}
where we again neglected inertial contributions and body forces and used the total stress $\bbar{\bm{\sigma}}_{\mathcal{B}^\pm}$ (see Eq.~\eqref{eq:barbar_sigmaB}), containing both mechanical and Maxwell contributions. Assuming the electrolytes behave like Newtonian fluids, the total stresses are \cite{fong2020transport}
\begin{align}
    \bbar{\bm{\sigma}}_{\mathcal{B}^\pm} = \underbrace{\vphantom{\biggl(} - p^\pm \bmI + \mu \left( \grad{\bmu^\pm} + \grad{\bmu^\pm}^T\right) + \chi \divv{\bmu^\pm}\bmI}_{\bm{\sigma}_{\mathcal{B}^\pm}} + \underbrace{\varepsilon_{\mathcal{B}}\left( \bme_{\mathcal{B}^\pm} \otimes \bme_{\mathcal{B}^\pm} - \frac{1}{2}  \bme_{\mathcal{B}^\pm} \cdot \bme_{\mathcal{B}^\pm} \bmI \right)}_{\bm{\sigma}_{\mathrm{M}\mathcal{B}^\pm}}~, \label{eq:bbarNewtMaxFull}
\end{align}
with the bulk pressure $p^\pm$, shear viscosity $\mu$, bulk viscosity $\chi$, and the electric fields $\bme_{\mathcal{B}^\pm}= -\grad{\phi_{\mathcal{B}^\pm}}$.

\paragraph{Membrane}

As in the bulk domain, we use the assumption of electrostatics and describe the membrane as a linear dielectric electric material. Then, the zeroth-order membrane potential $\phi_{0}$ satisfies 
\begin{align}
    \varepsilon_\mathcal{M} & \laplaces{\phi_{0}} - 4 C_\mathcal{M} \phi_{1} H +  \frac{16}{\delta} C_\mathcal{M} \phi_{2} = 0~,
\end{align}
and the first- and second-order membrane potentials are given by
\begin{align}
    \phi_1 &= -\frac{1}{2 C_\mathcal{M}} \left(\bm{n} \cdot \langle \varepsilon_\mathcal{B} \bm{e}_{\mathcal{B}} \rangle^\mathcal{M} - \frac{1}{2}\left( \sigma^+ - \sigma^- \right)\right)~,\\
    \phi_2 &=  -\frac{1}{16 C_\mathcal{M}} \left(\bm{n} \cdot \llbracket \varepsilon_\mathcal{B}\bm{e}_{\mathcal{B}} \rrbracket^\mathcal{M} - \left(\sigma^+ + \sigma^-\right)\right)~,
\end{align}
with the average $\langle \varepsilon_{\mathcal{B}} \bm{e}_{\mathcal{B}} \rangle^\mathcal{M} = \frac{\varepsilon_{\mathcal{B}}}{2}\left(\bm{e}_{\mathcal{B}^+}\rvert_{\mathcal{S}^+} + \bm{e}_{\mathcal{B}^-}\rvert_{\mathcal{S}^-} \right)$ and jump $\llbracket \varepsilon_{\mathcal{B}}\bm{e}_{\mathcal{B}} \rrbracket^\mathcal{M} = \varepsilon_{\mathcal{B}}\bm{e}_{\mathcal{B}^+}\rvert_{\mathcal{S}^+} - \varepsilon_{\mathcal{B}}\bm{e}_{\mathcal{B}^-}\rvert_{\mathcal{S}^-}$.\textspace

Given the common assumption that lipid membranes are incompressible, the membrane mass balance reduces to 
\begin{align}
    v^\alpha_{:\alpha } - 2 v H  = 0~.
\end{align}
In the absence of spontaneous curvatures, the in-plane components of the linear momentum balance for the incompressible case can be written as 
\begin{align}
     & 0 = \lambda_{,\Gb} a^{\Ga\Gb} + \frac{k_\mathrm{b}}{2} K_{,\Gb}a^{\Ga\Gb} + \pi^{\Gb\Ga}_{;\Gb} + \bm{f}_\mathrm{s} \cdot \bm{a}^\Ga + \left( \left(\bm{t}^+ + \bm{t}^-\right) + \left\llbracket \left\llbracket \bm{\sigma}^T_\mathrm{M} \right\rrbracket \right\rrbracket^\mathcal{M} \bm{n} \right) \cdot \bm{a}^\Ga \nonumber \\ 
    & \hspace{0.34\textwidth}  -\delta\left( 2H\delta_\Gg^\Ga + b_\Gg^\Ga\right)\left(\frac{1}{2}\left(\bm{t}^+ - \bm{t}^-\right) + \left\langle \left\llbracket \bm{\sigma}^T_\mathrm{M} \right\rrbracket \right\rangle^\mathcal{M} \bm{n}\right) \cdot \bm{a}^\Gg~, \label{eq:2pd_inplane_incomp_CG0}
\end{align}
and the shape equation is given by
\begin{align}
    &0 = 2
    \lambda H - k_\mathrm{b}\left(4H^2 - 3K\right)H - {k}_\mathrm{b}\Delta_\mathrm{s} H  + \pi^{\alpha\beta}b_{\alpha\beta} +   P^\Ga_{:\Ga} + \left( \bm{t}^+ + \bm{t}^- + \left\llbracket \left\llbracket \bm{\sigma}^T_\mathrm{M} \right\rrbracket \right\rrbracket^\mathcal{M} \bm{n}\right) \cdot \bm{n} \nonumber \\
    &  \hspace{0.385\linewidth} - 2\delta H \left(\frac{1}{2}\left(\bm{t}^+ - \bm{t}^-\right) + \left\langle \left\llbracket \bm{\sigma}^T_\mathrm{M} \right\rrbracket \right\rangle^\mathcal{M} \bm{n}\right)\cdot\bm{n} + \bm{f}_{\mathrm{s}} \cdot \bmn~, \label{eq:2pd_shape_incomp_CG0}
\end{align}
with the average $\left\langle \left\llbracket \bm{\sigma}^T_\mathrm{M} \right\rrbracket \right\rangle^\mathcal{M}$ and jump $\left\llbracket \left\llbracket \bm{\sigma}^T_\mathrm{M} \right\rrbracket \right\rrbracket^\mathcal{M}$ defined in Eqs.~\eqref{eq:sigM_avg} and \eqref{eq:sigM_diff}, respectively. Furthermore, we used the short-hand notation,
\begin{align}
    \pi^{\Ga\Gb} &= \zeta \left( w^\Ga_{:\Gg}a^{\Gg\Gb} + w^\Gb_{:\Gg} a^{\Gg\Ga} \right) + \bar{\omega} \left( v^\Ga_{0:\Ga} - 2vH \right)~,
\end{align}
for the membrane viscous stresses as well as
\begin{align}
    P^\Ga &= \delta\left( \frac{1}{2}\left(\bm{t}^+ - \bm{t}^-\right) + \left\langle \left\llbracket \bm{\sigma}^T_\mathrm{M} \right\rrbracket \right\rangle^\mathcal{M} \bm{n} - \frac{\delta H}{4}\left(\bm{t}^+ + \bm{t}^- + \frac{1}{2}\left\llbracket \left\llbracket \bm{\sigma}^T_\mathrm{M} \right\rrbracket \right\rrbracket^\mathcal{M} \bm{n}\right) \right)\cdot \bma^\Ga~.
\end{align}
The mechanical tractions $\bmt^\pm$ acting on the membrane from the bulk can be computed from the Newtonian material model assumed for the bulk fluid in Eq.~\eqref{eq:bbarNewtMaxFull} using $\bmt^\pm = \pm \bm{\sigma}_{\mathcal{B}^\pm}\rvert_{\mathcal{S}^\pm}\bmn$. 

\paragraph{Coupling conditions}
In applying the dimension reduction procedure to Gauss' law and the balance laws, we readily incorporated the electric field jump conditions and traction continuity conditions. Thus, we are only left with enforcing continuity of the electric potential,
\begin{align}
    \phi_0 \pm \phi_1 + \phi_2 - \phi_{\mathcal{B}^\pm}\rvert_{\mathcal{S}^\pm} &= 0~,\label{eq:couplingPhi}
\end{align}
no-flux conditions \cite{alkadri2024irreversiblethermodynamicscurvedlipid},
\begin{align}
    \bmJ_i \bigr\rvert_{\mathcal{S}^\pm} = \bm0~, \quad i = 1,..., N~, \label{eq:no-flux}
\end{align}
as well as velocity continuity,
\begin{align}
    \left(v_0^\Ga \pm \frac{\delta}{2}v_1^\Ga\right)\bma_\Ga + v_0^3 \bmn - \bmu^\pm\big\rvert_{\mathcal{S}^\pm}  = \bm0~, \label{eq:couplingV}
\end{align}
on the membrane-bulk interfaces $\mathcal{S}^\pm$. Note that the no-flux conditions in Eq.~\eqref{eq:no-flux} arise from the assumption that there are no chemical reactions and that the membrane is impermeable to solutes or solvent~\cite{alkadri2024irreversiblethermodynamicscurvedlipid}.

\paragraph{Membrane boundary conditions}
The electrostatics equations can be closed with the Dirichlet boundary condition
\begin{align}
    \phi_0 = \bar{\phi}_{\mathcal{M}0}
\end{align}
or Neumann boundary condition
\begin{align}
    -{\nu}^\alpha \left( \phi_{0,\alpha} + \frac{\delta}{4} \phi_{1,\beta} b^\beta_\alpha + \frac{\delta^2}{16} \phi_{2,\beta} b^\beta_\gamma b^\gamma_\alpha \right) = \bar{e}_0~,
\end{align}
or a combination thereof \cite{omar2024ES}, with $\bar{\phi}_{\mathcal{M}0}$ and $\bar{e}_0$ being the zeroth-order coefficients of the prescribed potential and normal component of the electric field, respectively. Note, however, that the potential must be prescribed at at least one point of either the membrane or bulk domain boundary.\textspace

Furthermore, either the Dirichlet boundary conditions
\begin{alignat}{2}
    v_0^\alpha &= \langle \bar{\bm{v}} \cdot \bm{a}^\alpha, \ChebT{0}{\Theta} \rangle~, \label{eq:BCSparD_inplane0_summary}  \\
    v_0^3 &= \langle \bar{\bm{v}} \cdot \bm{n}, \ChebT{0}{\Theta} \rangle~, \\
    \delta v_1^\alpha &= 4\langle \bar{\bm{v}} \cdot \bm{a}^\alpha, \ChebT{1}{\Theta} \rangle~,\label{eq:BCSparD_inplane1_summary}
\end{alignat}
or the Neumann boundary conditions 
\begin{align} 
    \check{\nu}^\Gb \left( \lambda + \frac{1}{2}k_\mathrm{b}K \right) + \check{\nu}_\Ga \pi^{\Ga\Gb} &= \delta\langle \bma^\Gb \cdot \tilde{\bar{\bm{t}}}, \ChebT{0}{\Theta}\rangle~,\label{eq:t0_inplane_BC_incomp_6}\\
    \check{\nu}_\alpha \left(k_\mathrm{b}Ha^{\Ga\Gb} + \frac{\delta^2}{8}b^{\Ga}_\Gg \pi^{\Gg\Gb} \right) &= -\frac{\delta^2}{2}\langle \bma^\Gb \cdot \tilde{\bar{\bm{t}}}, \ChebT{1}{\Theta}\rangle~. \label{eq:t1_inplane_BC_incomp_6} 
\end{align}
need to be enforced at every point on the boundary of the membrane. However, recall that only Eqs.~\eqref{eq:BCSparD_inplane0_summary}--\eqref{eq:BCSparD_inplane1_summary} or Eqs.~\eqref{eq:t0_inplane_BC_incomp_6} and \eqref{eq:t1_inplane_BC_incomp_6} can be enforced at any given point---but not a combination thereof---to maintain consistency with the three-dimensional perspective.

\section{Conclusion and Outlook}
This article concludes the derivation of a self-consistent surface theory of the electromechanics of deforming lipid membranes. In part 1, we proposed a new dimension reduction method for three-dimensional differential equations defined on thin bodies yielding \textit{effective} surface differential equations and applied it to derive an effective electrostatics surface theory. To couple electrostatic and mechanical effects, we then derived dimensionally reduced electromechanical balance laws for thin bodies in part 2~\cite{omar2023BL}. 
In this article, we specialized these results to capture the in-plane fluid and out-of-plane elastic behavior of lipid membranes by proposing suitable constitutive models and deriving the corresponding equations of motion and boundary conditions.
These equations, together with the electrostatics surface theory proposed in part 1, constitute the $(2+\delta)$-dimensional theory of the electromechanics of lipid membranes.\textspace

We refer to the proposed theory as ``$(2+\delta)$-dimensional'' to indicate that the governing equations are defined on a surface while also capturing effects associated with the thickness $\delta$. As argued in part 1 \cite{omar2024ES}, an electromechanical theory of lipid membranes cannot be directly formulated on a surface, as direct or strict surface theories cannot capture (i) potential differences across a membrane, (ii) the effects of distinct surface charges on the two interfaces between a membrane and its surrounding fluid, and (iii) the electric field within a membrane. These phenomena, in turn, affect the forces acting on lipid membranes and need to be captured in an electromechanical theory of lipid membranes, making it necessary to account for their finite thickness. 
The proposed dimension reduction method and the ensuing $(2+\delta)$-dimensional theory offer one such self-consistent construction. \textspace

The proposed $(2+\delta)$-dimensional theory opens up the study of a number of new applications. Among them is the electromechanical coupling arising from mechanically-gated ion channels such as TRAAK \cite{brohawn2014mechanosensitivity} and Piezo1 \cite{coste2010piezo1, cox2016removal}, whose openings are controlled by membrane tension. 
The flux of ions resulting from channel opening leads to changes in ion concentrations in the domains on either side of the membrane~\cite{row2024spatiotemporal}, inducing electromechanical forces in the membrane which, in turn, affect the opening probability of the channels. Such effects can only be addressed with a framework that consistently captures electromechanical effects, such as the $(2+\delta)$-dimensional theory described here. 
Many other applications such as deformations of lipid membranes under applied electric fields~\cite{winterhalter1988deformation, kummrow1991deformation,dimova2007giant, dimova2009vesicles, vlahovska2019electrohydrodynamics, portet2012destabilizing, needham1989electro, riske2005electro} are also worthy of future study. 

\section*{Acknowledgements}
The authors would like to thank Dr. Sirui Ning for carefully verifying all derivations provided in this article. This work is supported primarily by Director, Office of Science, Office of Basic Energy Sciences, of the U.S. Department of Energy under contract No. DEAc02-05CH11231. KKM also acknowledges support from the Hellman Foundation. ZL was supported partially by the National Science Foundation through NSF-DFG 2223407 and the Deutsche Forschungsgemeinschaft (German Research Foundation)—509322222. All authors acknowledge support to KKM from UC Berkeley.

	%
	%

	\newpage
	\addcontentsline{toc}{section}{References}
	\bibliographystyle{articleFiles/bibliography/bibStyleCap}
	\bibliography{articleFiles/bibliography/bibliography}

\end{document}